\documentclass[12pt]{article}
\pdfoutput=1 
\usepackage[utf8]{inputenc}
\usepackage[T1]{fontenc}
\usepackage{amsmath,amssymb, amsfonts}
\usepackage{icomma}
\usepackage[utf8]{inputenc}
\usepackage[T1]{fontenc}
\usepackage{color}
\usepackage{graphicx}

\usepackage{titlesec}
\usepackage{lmodern}
\usepackage{units}
\usepackage{marvosym}
\usepackage{MnSymbol}
\usepackage[usenames,dvipsnames,svgnames,table]{xcolor}
\usepackage{cite}
\usepackage{tikz}
\usepackage[small]{subfigure}

\newcommand{\sign}{\text{sign}}

\begin{document}

\makebox[0pt][l]{\hspace*{120mm} \parbox{3cm}{ICCUB-15-005}}
\hspace*{-16mm}

\renewcommand{\thefootnote}{\fnsymbol{footnote}}
\setcounter{footnote}{0}

\bigskip\bigskip

\begin{center}
{\Large\textbf{\mathversion{bold} 
 ABJM Theory with mass and FI deformations and Quantum Phase Transitions
}
\par}

\vspace{0.8cm}

 \textrm{Louise~Anderson$^{1}$ and
Jorge~G.~Russo$^{2,3}$}
\vspace{4mm}

\textit{${}^1$Department of Fundamental Physics\\
Chalmers University of Technology\\
S-412 96 G\"oteborg, Sweden}\\
\textit{${}^2$ Instituci\'o Catalana de Recerca i Estudis Avan\c cats (ICREA),\\
Pg. Lluis Companys 23, 08010 Barcelona, Spain.}\\
\textit{${}^3$ ECM Department and Institute of Cosmos Sciences, Facultat de F\'\i sica, \\
Universitat de Barcelona, 
Mart\'i Franqu\`es 1, E08028 Barcelona, Spain.}\\
\vspace{0.2cm}
\texttt{louise.anderson@chalmers.se, jorge.russo@icrea.cat}

\vspace{1em}
\begin{abstract}
 The phase structure of  ABJM theory with mass $m$ deformation and non-vanishing Fayet-Iliopoulos (FI) parameter, $\zeta$, is studied through the use of localisation on ${\mathbb S}^3$. The partition function of the theory then reduces to a matrix integral, which, in the large $N$ limit and at large sphere radius, is exactly computed by a saddle-point approximation. When the couplings are analytically continued to
real values, the phase diagram of the model becomes immensely rich, with an infinite series of third-order phase transitions  at vanishing FI-parameter \cite{Anderson:2014hxa}. As the  FI term is introduced, new effects appear. For any given $0<\zeta <m/2$, the number of phases is finite and for $\zeta\geq m/2$ 
the theory does not have any phase transitions at all.
 Finally, we argue that ABJM theory with physical couplings does not undergo phase transitions
and investigate the case of $U(2)\times U(2)$  gauge group in detail by an explicit calculation of the partition function.

\end{abstract}

\end{center}

\newpage

\tableofcontents
\newpage

\section{Introduction}

In some cases, in supersymmetric gauge theories, observables with enough supersymmetry may be computed exactly. For example, this applies to the partition function in ${\cal N}=2$ four-dimensional gauge theories on $S^4$ through the technique of localisation \cite{Pestun:2007rz}. This allows us to write observables with a sufficient amount of supersymmetry, such as the partition function, in terms of matrix integrals. Even though being much simpler than the original functional integrals, these matrix integrals still carry important information of the underlying field theory.

Under favourable conditions, these integrals may sometimes be determined exactly, though this often amounts to going to the large $N$, or planar, limit. In this limit, one may use techniques from random matrix theory to solve these matrix integrals \cite{Brezin:1977}. 
One of the most interesting applications of this approach is perhaps the test it provides of the conjectured gauge/gravity-duality: localisation and large $N$-techniques have provided insights 
into the strong-coupling behaviour of field theories with holographic duals, thus allowing for direct comparisons of quantities with a non-trivial dependence of the coupling constant between the gauge- and gravity-side of the duality, with excellent agreement
(for a review, see \cite{Marino:2011nm}).

In recent years,  the occurrence of quantum phase transitions in massive gauge theories in the  decompactification limit  (where the radius of the sphere is taken to infinity), has been found in a wide variety of supersymmetric theories. 
They were first found  for 
   $\mathcal{N}=2$ four-dimensional gauge theories with massive matter \cite{Russo:2013qaa}, whose critical properties were further investigated in 
\cite{Russo:2013kea,Russo:2013sba,Chen:2014vka,Marmiroli:2014ssa,Zarembo:2014ooa,Russo:2014nka}),
and, soon after, new examples were found
in three- \cite{Barranco:2014tla,Russo:2014bda,Anderson:2014hxa},  and five dimensions
\cite{Minahan:2014hwa}.
A  generic type of phase transition occurs when, at specific couplings, extra massless states
appear in the spectrum and contribute to the saddle-point. It is a resonance phenomenon,
as explained in \cite{Russo:2013qaa}. These resonance effects already appear in theories with  fundamental matter, but become much more complicated when the theory contains adjoint matter, 
leading to an infinite sequence of secondary resonances and a much richer phase structure. In four dimensions, the simplest example with massive adjoint matter is the $\mathcal{N}=2^*$ theory, which, in the decompactification limit, exhibits an infinite series of weak/strong quantum phase transitions as the coupling grows. Moreover, these phase transitions accumulate at strong coupling \cite{Russo:2013qaa,Russo:2013sba, Russo:2013kea}, raising interesting questions about how they manifest in the holographic dual theory. These were studied in detail in \cite{Chen:2014vka, Zarembo:2014ooa}, and similar phase transitions are also found to be present when the four-sphere is squashed \cite{Marmiroli:2014ssa}. 
Recently, it was shown that these non-trivial phases induce non-analytic dependence of
higher-rank Wilson loops on the rank parameter \cite{zarembonew}, which could in principle have a holographic counterpart in terms of  phase transitions in the effective field theory on the dual D brane description.

This type of phase transitions were also shown to occur for low-rank gauge groups, such as  $\mathcal{N}=2$ $SU(2)$ gauge theory with fundamental matter \cite{Russo:2014nka}, i.e.  supersymmetric $SU(2)$ QCD with two massive flavours. 
In this case, 
the quantum critical point of these phase transitions is identified 
with the Argyres-Douglas superconformal  fixed point of the theory. For $SU(2)$ gauge group, the phase transition
is driven by instantons (unlike the large $N$ case, where they are  negligible), and, in the decompactification limit,   the   free energy $\log Z$  including instantons can be expressed in terms of  the Seiberg-Witten prepotential  --a connection that arises when a saddle-point exists \cite{Russo:2014nka}. 

Similar phase transitions occur in three-dimensional Chern-Simons  theories with massive matter.
For example, three-dimensional Chern-Simons theory with massive fundamental matter has been shown to undergo  three distinct phases as the coupling runs from zero to infinity \cite{Barranco:2014tla}. These phase transitions have furthermore been investigated at finite $N$, where  the finite $N$ partition function of $U(N)$ Chern-Simons theory with massive fundamental matter was computed  using the theory of Mordell integrals  \cite{Russo:2014bda}.

The theory which will be in focus in this paper is the three-dimensional ABJM theory, which, in the massless case is an $\mathcal{N}=6$ superconformal theory dual to type IIA string theory on $AdS_4 \times CP^3$ \cite{Aharony:2008}. By using localisation techniques on the field theory side, the partition function can  be computed exactly at large $N$ \cite{Kapustin:2009kz, Drukker:2010nc}, which may then be compared to geometric analyses on the string theory side, enabling extensive tests of the gauge/gravity conjecture \cite{Drukker:2010nc, Herzog:2010hf}. 
The partition function can be computed exactly
 even in the mass-deformed case as the path integral localises to a matrix model on ${\mathbb S}^3$ \cite{Kapustin:2009kz,Kapustin:2010xq}.
Recently, it was shown that a version of ABJM theory
obtained by analytic continuation in the couplings exhibits phase transitions  \cite{Anderson:2014hxa}.
This theory contains bi-fundamental matter and the resulting phase structure
 arising in the  decompactification limit resembles the case of  four-dimensional $\mathcal{N}=2^*$ theory, with an infinite number of third-order phase transitions accumulating at strong coupling. 
These models corresponds to computing the partition function in the region of parameter space where
the couplings are real (and the Chern-Simons levels are imaginary). This method, where one starts with  unphysical couplings, was successfully implemented in the past \cite{Marino:2011nm,Drukker:2010nc}
to compute the large $N$ partition function for the ABJM theory with no mass deformation.
A direct, analytic calculation of the large $N$ partition function in ABJM theory with physical Chern-Simons levels
is more complicated and has not been carried out so far. The reason is that eigenvalues appear to be distributed in different cuts in the complex plane with non-homogeneous $N$ dependence
for real and imaginary parts (see e.g. a discussion in \cite{Herzog:2010hf}).

In \cite{Anderson:2014hxa}, the Fayet-Iliopoulos (FI) parameter was set to zero. The aim of this work is to incorporate this parameter, which, as we shall see, significantly enriches the phase structure of the theory. 
 As the FI-parameter approaches zero, the phase structure found herein agrees with previous results, and furthermore serves to clarify some of the peculiar behaviour previously seen. However for non-vanishing FI-parameter, the situation is, as mentioned, significantly different, and the solutions to the saddle-point equations are divided into cases depending on the precise relations between the mass-deformation parameter $m$ and the FI-parameter, $\zeta$. The most dramatic effect is that the number of phase transitions undergone as the coupling is increased are now finite, and  for $\zeta>m/2$
there is only a trivial phase with constant eigenvalue density in some finite region of support.

The structure of this work is as follows: In section  \ref{sec:ABJM}, we give the matrix model representation of the partition function for ABJ(M) theory with mass and FI deformation parameters, and discuss some basic properties.
In section \ref{nuevasec}, the analytical continuation  is introduced. In  subsection \ref{sec:eqqq}, we specialise to the situation where the couplings $\lambda_1,\ \lambda_2$ for the two (analytically continued) gauge groups are equal and real, and in  sections  \ref{sec:large_FI} and \ref{sec:small_FI}, we determine the eigenvalue densities for the theory in the large $N$ limit. As mentioned, the precise form of these depend on the relations between $m$ and $\zeta$, and these different cases are considered in sections  \ref{sec:large_FI} and \ref{sec:small_FI}.
 In section \ref{sec:Conclusion},  we provide a summary of our results for the cases of equal and real couplings, 
and also discuss the case of $\lambda_1 \neq \lambda_2$ ; $ \lambda_{1,2} >0$, for the two different schemes of analytic continuations.
In section \ref{abjmtwo}, we present a general argument showing that physical ABJM theory
does not have phase transitions. The absence of phase transitions is illustrated by
considering the case with $U(2)\times U(2)$ gauge group, where 
 the partition function can be explicitly computed both for the analytically continued model and for the model with physical couplings. The analytically continued model exhibits analogous phase transitions as in the large $N$ case. However, the physical ABJM theory does not undergo any phase transition.
One key feature that seems to make physical ABJM theory to be non-generic is  the fact that the  Chern-Simons levels for the gauge groups are equal and opposite.
 Finally, in section \ref{concluding}, we end with some concluding remarks. Details of the calculation of section   \ref{sec:mtilde_geq_B-A}  are given in an appendix.


\section{Deformed ABJ(M) theory \label{sec:ABJM}}


 ABJM theory is a three-dimensional superconformal theory with maximal supersymmetry, gauge group $U_k(N) \times U_{-k}(N)$, and matter in the bifundamental representation. $k$ denotes the Chern-Simons level of the two gauge groups respectively. In ABJ theory \cite{Aharony:2008gk}, the situation is  generalised so that the two gauge groups are allowed to have diffterent ranks. Allowing for this small generalisation was shown to be useful when computing the partition function of the massless theory by analytic continuation  \cite{Drukker:2010nc},
as well as in the mass-deformed case \cite{Anderson:2014hxa}. 

In the massless case, these theories are well-studied and their path integrals on ${\mathbb S}^3$ are known to localise onto constant field configurations \cite{Kapustin:2009kz, Jafferis:2010un}.
 However, these results do not rely on the theory being conformal, nor maximally supersymmetric, and thus they may be used to examine the deformed ABJ theory as well. 

In previous work, the decompactification limit of the mass-deformed ABJ theory was considered \cite{Anderson:2014hxa}, and quantum weak/strong phase transitions were shown to be present in two analytically continued versions of the theory.

However, there is another way of introducing a scale to the problem, other than introducing a mass: by introducing a Fayet-Illiopoulos deformation. 
This will be present in the most general form of supersymmetric deformation, and it  is this setup which will be considered herein, with both non-vanishing mass and FI-parameter. The partition function on ${\mathbb S}^3$ takes the form of an eigenvalue integral in the large $N$-limit \cite{Kapustin:2009kz,Kapustin:2010xq}, and, with the normalisation of \cite{Drukker:2010nc}, may be written as:
 \begin{align}
 \label{eq:matrix_model_FI}
 Z_{\rm ABJM}(2\zeta, m; k) =& \; \frac{1}{N_1!\,N_2!} \int \prod_{i=1}^{N_1}\frac{ d\mu_i}{2 \pi}\,\,\prod_{a=1}^{N_2} \frac{d\nu_a}{2 \pi}\,\, 
 \\ \nonumber & 
 \times 
 \frac{\prod\limits_{i< j} \sinh^2\frac{\mu_i -\mu_j}{2} 
 \prod\limits_{a< b} \sinh^2\frac{\nu_a -\nu_b}{2}}
 {\prod\limits_{i\, a} \cosh\frac{\mu_i-\nu_a+m}{2}
 \cosh\frac{\mu_i-\nu_a-m}{2}
  }\,\,
 \,{\rm e}\,^{- \frac{i k}{2\pi} \zeta \left(
 \sum\limits_{i}^{}\mu _i+\sum\limits_{a}^{}\nu _a
 \right) + 
 \frac{ik }{4\pi } \left(
 \sum\limits_i \mu_i^2-\sum\limits_{a}^{}\nu_a^2
 \right)
 }
 ,\end{align}
where $\mu_i,\nu_a$ represent the eigenvalues of the auxiliary fields from the vector multiplets of the two gauge groups, $m$ denotes the mass-deformation
and $\zeta $ represents the FI-parameter, which will be taken to be real.

By shifting the integration variables, $x \equiv \mu - \zeta, y\equiv \nu+\zeta$, we may then move the FI-dependence from the exponent to the denominator, and what we are left with may be thought of as the mass-deformed case but with \emph{two different masses}, $m_1$ and $m_2$.  In these variables, the partition function becomes:
 \begin{align}
 \label{eq:matrix_model_FI_shifted}
 Z_{\rm ABJM}(2\zeta, m; k) =& \; \frac{{\rm e}^{\frac{\zeta^2(N_2-N_1)}{2g}}}{N_1!\,N_2!} \int \prod_{i=1}^{N_1}\frac{ dx_i}{2 \pi}\,\,\prod_{a=1}^{N_2} \frac{dy_a}{2 \pi}\,\, 
 \\ \nonumber & 
 \times 
  \frac{ \prod \limits_{i < j} \sinh^2\frac{x_i -x_j}{2} 
 \prod \limits_{a< b} \sinh^2\frac{y_a -y_b}{2}}
 {\prod \limits_{i, a} \cosh\frac{x_i-y_a+m_1}{2} 
 \cosh\frac{x_i-y_a-m_2}{2}
  }\,\,
 \,{\rm e}\,^{- 
 \frac{1 }{2 g } \left(
 \sum\limits_i x_i^2-\sum\limits_{a}^{}y_a^2
 \right)
 }
, \end{align}
where $g=\frac{2 \pi i}{k}$ represents the coupling, and
$m_1,m_2$ relates to $m$ and $\zeta$ as:
\begin{align}
m_1=m+2\zeta  \hspace*{6mm} \text{and} 
\hspace*{6mm} m_2=m-2\zeta\ .
\end{align}

Consider the partition function \eqref{eq:matrix_model_FI} with $N_1=N_2\equiv N$.
It can be written in another  form, which is useful to exhibit some symmetries. This is done by following \cite{Kapustin:2010xq}, slightly generalising their derivation
for $k=1$ to arbitrary $k$. By using the identity,
\begin{align}
\frac{\prod_{i<j} \sinh(x_i-x_j)\sinh(y_i-y_j) }{\prod_{i,j} \cosh(x_i-y_j) } =\sum_\rho
(-1)^\rho \prod_i \frac{1}{\cosh(x_i-y_{\rho(i)})}\ ,
\end{align}
where $\rho $ runs over all permutations of $\{ 1,...,\ N\} $, the partition function may be written as:
\begin{align}
Z_{\rm ABJM}(2\zeta, m; k) =&  \sum_{\rho, \rho'} \frac{(-1)^{\rho+\rho'}}{N!^2}
\int \frac{ d^N\mu}{(2 \pi)^N}\,\ \frac{d^N\nu}{(2 \pi)^N}\, \prod_i\, 
 \frac{    {\rm e}\,^{- \frac{i k}{2\pi} \zeta (\mu _i +\nu_i)+ 
 \frac{ik }{4\pi } (\mu _i^2 -\nu_i^2) }}
 { \cosh\frac{\mu_i-\nu_{\rho(i)}+m}{2} \cosh\frac{\mu_i-\nu_{\rho'(i)}-m}{2}
  } 
\nonumber \\
&= \sum_{\rho} \frac{(-1)^{\rho}}{N!}
\int \frac{ d^N\mu}{(2 \pi)^N}\,\ \frac{d^N\nu}{(2 \pi)^N}\, \prod_i\, 
 \frac{    {\rm e}\,^{- \frac{i k}{2\pi} \zeta (\mu _i +\nu_i)+ 
 \frac{ik }{4\pi } (\mu _i^2 -\nu_i^2) }}
 { \cosh\frac{\mu_i-\nu_i+m}{2} \cosh\frac{\mu_i-\nu_{\rho(i)}-m}{2}
  } \ .
\end{align}
We now make use of the Fourier transform
\begin{align}
\label{fouri}
\int d\tau  \frac{{\rm e}\,^{ i \tau \mu}}{\cosh \pi \tau} =\frac{1}{\cosh  \frac{\mu}{2} } \ ,
\end{align}
for all hyperbolic cosines in the denominator, introducing new integration variables $\tau_i,\ \tau_i'$. The integrals over $\mu_i, \ \nu_i$ then become Gaussian and can be computed  explicitly, after which one finds
\begin{align}
\label{jain}
 Z_{\rm ABJM}(2\zeta, m; k) = \sum_{\rho} \frac{(-1)^{\rho}}{k^N N!} \int d^N\tau d^N \tau'
\frac{ {\rm e}\,^
{-\frac{2\pi i}{k} \sum_i \tau_i '(\tau_i -\tau_{\rho(i)})  +i\sum_i ( \tau_i' m_1 -\tau_i m_2) }}{
\prod_i \cosh(\pi \tau_i ) \cosh(\pi \tau'_i)} \ .
\end{align}
Using again the Fourier transform (\ref{fouri}) and computing  the integral over $\tau'_i $, we finally obtain (upon rescaling $\tau_i\to k\tau_i $)
\begin{align}
\label{permuform}
Z_{\rm ABJM}(2\zeta, m; k)=\sum_\rho (-1)^\rho \frac{1}{N!} \int d^N\tau \frac{ e^{ -i k m_2 \sum_i\tau_i} }{\prod_i \cosh( k \pi \tau_i) \cosh(\pi(\tau_i -\tau_{\rho(i)}) -\frac{m_1}{2}) } .
\end{align}

We stress that the derivation  above only holds when the Chern-Simons levels of the two gauge groups $U(N)_{k_1}\times U(N)_{k_2}$
are  opposite, $k_2=-k_1$, which is the case in ABJM theory 
(for $k_2\neq -k_1$, terms $\tau_i'{}^2, \tau_i^2$ remain in the exponent, leading to more complicated expressions).

The partition function \eqref{eq:matrix_model_FI} has the obvious symmetry $\zeta \to -\zeta $,
under which $m_1  \leftrightarrow m_2$. However,  the partition function written in the form \eqref{permuform}  makes manifest another symmetry,
\begin{align}
m_2\leftrightarrow -m_2\ ,
\end{align}
arising after the sum over permutations.
Under this symmetry, the FI- and mass-deformations are exchanged.
In other words,  the deformed ABJM partition function \eqref{eq:matrix_model_FI} with $N_1=N_2$ 
enjoys the property
\begin{align}
\label{fich}
Z_{\rm ABJM}(2\zeta, m; k)  = Z_{\rm ABJM}(m,2\zeta; k) \ .
\end{align}
In particular, a FI-deformation on the massless theory $\zeta=m/2$  is equivalent to a mass-deformation $m$ in the theory with vanishing FI-parameter,
\begin{align}
\label{fizz}
Z_{\rm ABJM}(m, 0; k)  = Z_{\rm ABJM}(0,m; k) \ .
\end{align}

\section{Analytic continuation and saddle-point equations \label{nuevasec}}

Following \cite{Anderson:2014hxa}, we first assume independent
Chern-Simons levels $k_1$ and $k_2$ for the two gauge groups $U(N_1)$ and $U(N_2)$, and introduce two different couplings, 
\begin{equation}
g_1= \frac{2 \pi i}{k_1}\ ,\qquad g_2= \frac{2 \pi i}{k_2}\ ,
\end{equation}
and equivalently, the two 't Hooft couplings 
\begin{equation}
\lambda_1 =N_1 g_1\ ,\qquad \lambda_2 =N_2 g_2\ ,
\end{equation}
for the different gauge groups.

Our starting point will be the representation \eqref{eq:matrix_model_FI_shifted} for the partition function.
In \cite{Anderson:2014hxa}, two analytic continuations of this model to arbitrary $\lambda_1,\ \lambda_2$ were considered: one where  we set $k_2=-k_1=k$ and leave $N_1,\ N_2$ arbitrary, which may be thought of as an analytic continuation in the gauge group rank. The other one may  be thought of as a continuation in the Chern-Simons level instead (and holds the rank of the two gauge groups equal). 
We will mainly use  the first analytic continuation (which is the one used in \cite{Drukker:2010nc,Marino:2011nm} for  ABJM theory at large $N$).
 The second analytic continuation will
be discussed in sections \ref{remak},  \ref{segunda} and later in section  \ref{Utwo}  for the special case where the gauge group is $U(2)\times U(2)$.

It is important to note that, after analytic continuation to two independent couplings $\lambda_1,\ \lambda_2$,
the resulting partition function $\hat Z(2\zeta,m;\lambda_1,\lambda_2) $ cannot be written  in  the form \eqref{permuform}, except in the special
case $\lambda_2=-\lambda_1$. In particular,  for generic  $\lambda_1,\ \lambda_2$,
the partition function does not satisfy the symmetry \eqref{fich}.

The ABJM partition function  \eqref{eq:matrix_model_FI_shifted} with integer $k$ is given by a {\it convergent} integral, therefore in principle one does not need to resort to analytic continuation
to define it. However, for integer $k$
 (thus imaginary $\lambda_2=-\lambda_1$), the saddle-points  lie in cuts in the complex plane which 
are complicated to determine even numerically. Here we perform analytic continuation to real, positive  couplings $\lambda_1, \lambda_2$ because it is in this case that the saddle-point equations can be solved explicitly in terms of closed formulas. 

With the analytic continuation in the gauge group rank, we set $k_2=-k_1=k$, and the saddle-point equations of \eqref{eq:matrix_model_FI_shifted} take the form:
\begin{align}
\label{eq:saddlepoints_FI}
x_i =
 &
 \frac{\lambda_1}{N_1} \sum_{ j \neq i}^{N_1} \coth\frac{x_i -x_j}{2} 
+    \frac{\lambda_2}{2 N_2} \sum_{a}^{N_2} \Big(  
 \tanh\frac{x_i-y_a+m_1}{2}  
 +\tanh\frac{x_i-y_a-m_2}{2}  
 \Big) \\ \nonumber 
y_a =
 &
 \frac{\lambda_2}{N_2}  \sum_{ b\neq a}^{N_2} \coth\frac{y_a -y_b}{2}  
+ \frac{\lambda_1}{2 N_1} \sum_{i}^{N_1} \Big(
 \tanh\frac{y_a-x_i- m_1}{2}
  + \tanh\frac{y_a-x_i+ m_2}{2}
 \Big) 
.\end{align}
The ABJM theory is recovered by analytic continuation $\lambda_1\to e^{i\varphi}\lambda $, 
 $\lambda_2\to e^{-i\varphi}\lambda $, where $\varphi$ goes from 0 to $\pi/2$. 

Phase transitions typically occur in the decompactification limit, where the radius $R$ of the three-sphere (set to unity in previous formulas) is sent to infinity. The dependence on the radius can be restored
by rescaling $m\to mR,\ \zeta\to \zeta R, \ x_i\to x_i R,\ y_i\to y_i R$. For the coupling, we take the same scaling used in \cite{Barranco:2014tla,Anderson:2014hxa}, where $\lambda/R$ is fixed as $R\to\infty$. This particular decompactification limit turns out to be self-consistent and the dependence on $R$ completely cancels from the saddle-point equations. 
As $R\to\infty$, the hyperbolic functions are replaced by sign functions. 
Furthermore, in the large $N$ limit, the eigenvalues $x_i,\ y_a$ have continuum distributions described by unit-normalised eigenvalue densities, $\rho_x (x ),\  \rho_y (y )$, and
the saddle-point equations  take the form
\begin{align}
\label{eq:saddlepoints_cont}
x =
 &
   \lambda_1 \int_{\mathcal{C}_x} dx'\,\rho _x (x ') \sign(x -x') 
   +   \frac{\lambda_2}{2}  \int_{\mathcal{C}_y} d y\,\rho _y (y ) \Big( \sign(x-y+m_1) + \sign(x-y-m_2) \Big) ,
 \\ \nonumber 
y =
 &
 \lambda_2  \int_{\mathcal{C}_y} d y' \,\rho_y (y ')\sign (y -y')  
+  \frac{\lambda_1}{2} \int_{\mathcal{C}_x} dx \,\rho_x (x ) \Big( \sign( y-x-m_1) +  \sign( y-x+m_2)\Big)
,\end{align}
where 
$\mathcal{C}_x$ and $\mathcal{C}_y$ denote the intervals on which $\rho_x$ and $\rho_y$ are supported respectively.
 
It will be convenient to introduce another change of variables, namely the rescaling $y \rightarrow -y$. By also defining $\hat{\rho}_y(y)=\rho_y(-y)$, such a rescaling leads to equations which are symmetric under the exchange of $x\leftrightarrow y$ (and as such also $\rho_x \leftrightarrow \hat{\rho}_y$ and the integration regimes) together with $\lambda_1 \leftrightarrow \lambda_2$.
These equations may explicitly be written down as:
\begin{align}
\label{eq:saddlepoints_cont_rescaled}
x =
 &
   \lambda_1 \int_{\mathcal{C}_x} dx'\,\rho _x (x ') \sign(x -x') 
   +   \frac{\lambda_2}{2}  \int_{\mathcal{C}_y} d y\,\hat{\rho} _y (y ) \Big( \sign(x+y+m_1) + \sign(x+y-m_2) \Big) ,
 \\ \nonumber 
y =
 &
\lambda_2  \int_{\mathcal{C}_y} d y' \,\hat{\rho} _y(y') \sign (y -y')  
+  \frac{\lambda_1}{2} \int_{\mathcal{C}_x} dx \,\rho_x (x ) \Big( \sign( y+x+m_1) +  \sign( y+x-m_2)\Big)
.\end{align}
Take $\mathcal{C}_x = [-A,B]$ and similarly $\mathcal{C}_y = [-C, D]$, where $\{ A,B,C,D\} \in \mathbb{R}$. 
Differentiating \eqref{eq:saddlepoints_cont_rescaled}  with respect to $x,y$ respectively gives us:
\begin{align}
\label{eq:saddles_ana1}
\rho_x(z) =& \frac{1}{2 \lambda_1}- \frac{\lambda_2}{2\lambda_1} \Big( 
\hat{\rho}_y(-z-m_1)+\hat{\rho}_y(-z+m_2)
\Big) ,
\\ \nonumber 
\hat{\rho}_y(z) =& \frac{1}{2 \lambda_2}- \frac{\lambda_1}{2\lambda_2} \Big(
\rho_x(-z-m_1)+\rho_x(-z+m_2)
\Big)
.\end{align}
Solving these coupled functional equations is very complicated in the general case.
For simplicity, in most of our discussion, the situation of \emph{equal, real couplings} will be considered.
In section \ref{diflam}, 
we will also treat the case of generic $\lambda_{1,2} >0$ and show that it exhibits the same qualitative features.

\subsection{The case of equal, real couplings  \label{sec:eqqq}}

By symmetry, it is clear from the expression of \eqref{eq:saddles_ana1} that, for $\lambda_1=\lambda_2=\lambda$, the system admits a solution with two equal densities $\rho_x, \hat{\rho}_y$. The problem thus reduces to finding the solution to one single equation for a density $\rho(z)$:
\begin{align}
\label{eq:saddles_ana1_equal}
\rho(z) =& \frac{1}{2 \lambda}-
 \frac{1}{2} \rho(-z-m_1)
- \frac{1}{2} \rho(-z+m_2)
,\end{align}
where $\rho(z)$ is supported on some interval $[-A,B]$ along the real axis. 

However, unlike the case  previously considered where the FI-parameter vanishes \cite{Anderson:2014hxa}, there is no reflection symmetry around the origin of these equations, and we cannot assume $\rho(-z)=\rho(z)$. This  complicates the situation  compared to the case studied in \cite{Anderson:2014hxa}.
With no loss of generality one can take  $\zeta>0$ and $m>0$.
Then $m_1$ will always be greater than zero whereas $m_2 \in [-\infty, m]$.

It is clear that the solution to this equation will behave qualitatively different depending on the sign of $m_2$, and we may thus divide our investigation into two separate cases:
\begin{itemize}
\item $m_1 > m_2,\ m_2 \leq 0$  corresponding to $\zeta \geq \frac{m}{2}$
\item $m_1 >0$, $m_2>0$, corresponding to $\zeta < \frac{m}{2}$
.\end{itemize}
These will be considered in sections \ref{sec:large_FI} and \ref{sec:small_FI} respectively.

\subsection{Phase transitions in Chern-Simons theory  with massive adjoint matter \label{remak}}

The saddle-point equations for the second analytic continuation may  be expressed in terms of $g_1,g_2$ and $N$, and will in this notation differ from the equations  \eqref{eq:saddlepoints_cont} of the first analytic continuation by  some signs. Defining $\alpha_1=Ng_1, \alpha_2=N g_2$, these equations may be written as:
\begin{align}
\label{eq:saddlepoints_cont2}
\frac{x}{\alpha_1} =
 &
  \int_{\mathcal{C}_x} dx'\,\rho _x (x ') \sign(x -x') 
   -   \frac{1}{2}  \int_{\mathcal{C}_y} d y\,\rho _y (y ) \Big( \sign(x-y+m_1) + \sign(x-y-m_2) \Big)
 \\ \nonumber 
\frac{y}{\alpha_2} =
 &
  \int_{\mathcal{C}_y} d y' \,\rho_y (y ')\sign (y -y')  
-  \frac{1}{2} \int_{\mathcal{C}_x} dx \,\rho_x (x ) \Big( \sign( y-x-m_1) +  \sign( y-x+m_2)\Big)
.\end{align}
ABJM theory is recovered by  analytically continuing $\alpha_1\to e^{i\varphi}\alpha_1$, $\alpha_2\to e^{-i\varphi}\alpha_1$, with $\varphi$ varying between $0$ and $\pi/2$ .

Consider the particular case $\alpha_1=\alpha_2\equiv \alpha $.
We are led to a single equation:
\begin{align}
\label{eq:saddles_adj}
\rho(x) =& \frac{1}{2 \alpha} + \frac{1}{2} \rho(-x-m_1)
+ \frac{1}{2} \rho(-x+m_2)
.\end{align}
If we further assume that $m_1=m_2\equiv m$ (i.e. $\zeta=0$), we have reflection symmetry, and the
equation becomes
\begin{align}
\label{eq:saddles_adj2}
\rho(x) =& \frac{1}{2 \alpha} + \frac{1}{2} \rho(x-m)
+ \frac{1}{2} \rho(x+m)
\ .\end{align}
The solutions to this equation were studied in section 4 of \cite{Anderson:2014hxa}.

It is interesting to compare this equation with the saddle-point equation that arises in a closely related system,
 $\mathcal{N}=3$ supersymmetric $U(N)$ Chern-Simons gauge theory with
two massive adjoint  multiplets. This is a precise three-dimensional analog of the $\mathcal{N}=2^*$ theory whose critical properties were
studied in \cite{Russo:2013qaa,Russo:2013kea,Russo:2013sba,Chen:2014vka,Zarembo:2014ooa}.
The partition function can be constructed with the general rules given in  \cite{Kapustin:2009kz}, whereupon one obtains:
 \begin{align}
 \label{adjoint}
 Z =& \; \frac{1}{N!} \int \prod_{i=1}^{N}\frac{ d\mu_i}{2 \pi}\, \prod_{i< j}
\frac{ \sinh^2\frac{\mu_i -\mu_j}{2} }
 {\cosh\frac{\mu_i-\mu_j+m}{2}  \cosh\frac{\mu_i-\mu_j-m}{2}}
 \,{\rm e}\,^{ - \frac{1 }{2 g } \sum\limits_i \mu_i^2 }\ .
 \end{align}
At large $N$ (and in the decompactification limit), this partition function can be computed by solving the same saddle-point equation 
\eqref{eq:saddles_adj2}, with $\alpha =g N$.
Therefore, the results of \cite{Anderson:2014hxa} equally apply to this case, and they can be summarized as follows: The theory has an infinite sequence of phase transitions, where in each phase  the eigenvalue density is  given  by a piecewise constant function. The number of discontinuities increases whenever the coupling crosses critical values
taking the theory into a new phase. 
In the strong   coupling limit, $\alpha >>1$,  the phase transitions accumulate and the equation becomes differential, $-m^2\rho''=1/\alpha$ and $\rho (x)$ approaches the  asymptotic form \cite{Anderson:2014hxa}
\begin{align}
\rho_\infty(x) = \frac{1}{2gm^2}(\mu_0^2-\mu^2)\ ,\qquad 
\mu_0= \left( \frac{3gm^2}{2}\right)^{\frac{1}{3}}\ .
\end{align}
In the infinite $\alpha $ limit, this smooth, parabolic asymptotic density arises as the envelope of (discontinuous) piecewise constant densities. 
In the case of $\mathcal{N}=2^*$ theory, at strong coupling, the eigenvalue density also reaches a smooth asymptotic form after going
through an infinite number of phase transitions, each phase described by a discontinuous density. In this case, the asymptotic density has the
 Wigner's semicircle shape, a property that has been  matched with the holographic prediction \cite{Buchel:2000cn,Buchel:2013id}.

In conclusion, $\mathcal{N}=3$   supersymmetric Chern-Simons gauge theory with
two massive adjoint multiplets has large $N$ phase transitions, with a  behaviour that shares similar features as its four-dimensional relative, $\mathcal{N}=2^*$ theory.


\section{Equal real couplings with $\zeta \geq \frac{m}{2}$ \label{sec:large_FI}}


Let us now move back to the problem of ABJM theory analytically continued in the gauge group rank in the special case of equal and real couplings for the two gauge groups, as described by equation \eqref{eq:saddles_ana1_equal}. Furthermore, let us specialise to the case of $\zeta \geq \frac{m}{2}$, corresponding to $m_1>0$, whereas $m_2 \leq 0$.


Let us first consider two limiting cases, namely $\zeta \rightarrow 0$ and $\zeta \rightarrow \infty$. 
In the first one of these, the situation reduces to the one considered in \cite{Anderson:2014hxa}, and equation \eqref{eq:saddles_ana1_equal} becomes symmetrical under reflection through the origin, implying $A=B$. 
As $\zeta$ goes to infinity,  equation \eqref{eq:saddles_ana1_equal} reduces to $\rho(z)=\frac{1}{2 \lambda}$, since the two shifted terms will vanish as their arguments will lie outside the region of support for $\rho$. Knowing the eigenvalue density, the two interval endpoints may be obtained from the integral equation \eqref{eq:saddlepoints_cont_rescaled} with $x=B$, together with the normalisation condition. From these, we find $A=0, B=2 \lambda$ for $\zeta \rightarrow \infty$ . 
It is then natural to  assume  that $B  \geq A $ at intermediate values (this is also confirmed by the numerical solution). 

With this assumption, together with the condition $m_1>-m_2>0$, the integral equation \eqref{eq:saddlepoints_cont_rescaled} with $x=B$  gives 
\begin{align}
B=2 \lambda \hspace*{3cm} \forall \zeta \geq \frac{m}{2}
.\end{align}
It can then be shown that the other interval endpoint  will always lie in the origin, i.e. $A=0$.

This gives us  an eigenvalue density as
\begin{align}
\rho(z)= \frac{1}{2 \lambda} \hspace*{3cm} \forall z \in [0, 2 \lambda]\ .
\end{align}
Therefore, there are  no phase transitions in this regime where $\zeta \geq \frac{m}{2}$.

\medskip

In conclusion, turning on a FI-parameter $\zeta \geq \frac{m}{2}$ implies a theory free
from phase transitions. {\it In particular, the masslesss theory with only FI deformation does not have
phase transitions.}

 
 \section{Equal real couplings with $\zeta < \frac{m}{2}$ \label{sec:small_FI}}


As the FI-parameter decreases below $m/2$, however, the situation becomes more complicated: $m_2$  changes sign, becoming strictly positive. 
When $\zeta=0$, one has $m_1=m_2$, which is precisely the case discussed in \cite{Anderson:2014hxa}. As previously mentioned, the saddle-point equation \eqref{eq:saddles_ana1_equal}  then has reflection symmetry, which gives us that $A=B$, i.e., the eigenvalue density is supported on the interval $[-B,B]$. 
On the other hand, as shown above, for all $\zeta \geq \frac{m}{2}$, the leftmost interval endpoint  lies at the origin.
As $\zeta$ increases from $0$ to $\frac{m}{2}$, the leftmost interval endpoint, $-A$,  thus moves from $-B$ to the origin. 

Precisely how this happens will depend on the coupling. 
We will start by considering an example, after which we proceed to the general solution.


\subsection{Simple examples
 \label{sec:special_case:n-0}}

Consider once again the saddle-point equation \eqref{eq:saddles_ana1_equal}.
To start with, assume that $m_1$ and $m_2$  are sufficiently large so that the points $-z-m_1$ and $m_2-z$
lie outside of the interval $[-A,B]$ (i.e. small $\zeta$), where the eigenvalue density has support.
This requires, in particular, that $m_2>2B$ and $m_1>2A$.
In this case the saddle-point equation reduces to
\begin{align}
\label{trivialcase}
\rho(z) =& \frac{1}{2 \lambda}
.\end{align}
Normalisation then gives the condition $A+B=2\lambda $.  The final condition arises from
the integral equation, giving $A=B$, hence $A=B=\lambda$.
This solution exists for a coupling where  the conditions $m_2>2B$ and $m_1>2A$ are satisfied,
i.e.  $0<\lambda< \frac{m_2}{2} $, since $m_2<m_1$.
For a larger $\lambda $, the second shifted term begins to contribute in some interval, and the
solution must therefore  experience a discontinuous change.

In general,  transitions occur when one of the two shifted terms,
$\rho(-z+m_2)$ or $\rho(-z-m_1)$, are turned on. This happens when a new {\it resonance point} , 
 reaches the interior of  $[-A,B]$.  Then,
$-z+m_2$ or $-z-m_1$ coincide with $-A$ or $B$. In terms of the variable $-z$, the  resonance points
are points located at a distance $m_2$ or $m_1$ from the endpoints of the interval.
Physically, the phase transitions occur because  beyond some critical couplings, extra massless particles (of masses proportional to $|-z+m_2|$ or to $|-z-m_1|$) begin to  contribute
to the partition function.

To proceed, we may consider the situation where $m_1>2A$,  which ensures that the first one of the shifted terms vanishes. The saddle-point equation \eqref{eq:saddles_ana1_equal} then reduces to 
\begin{align}
\rho(z) =& \frac{1}{2 \lambda}- \frac{1}{2} \rho(m_2-z)
,\end{align}
where we assume that $\rho(z)$ is supported on the interval $[-A,B]$, for some $0 < A< B$. 
The resonance point of the leftmost interval endpoint, $-A$, 
is then given by:
\begin{align}
a_1=m_2+A
.\end{align}
Similarly, the resonance of the interval endpoint $B$ is given by 
\begin{align}
\label{eq_res_B}
b_1=m_2-B,
\end{align}
 but these two points will simultaneously lie inside the interior of the interval $[-A,B]$ only in one limiting case. The extremal case to have a resonance originating from $-A$ corresponds to this resonance point coinciding with $B$, giving us a condition on $m_2$ as $m_2=B-A$.  When this condition is fulfilled, the resonance originating from $B$ will be $b_1=-A$.
 
  Thus, for $m_2 < B-A$, there will only be one resonance originating from the leftmost endpoint $-A$, and for  $2B> m_2 > B-A$, there will be one resonance originating from the rightmost endpoint $B$. 
 [As discussed above, for $m_2 > 2 B$, there are no resonance points inside the interval, and the solution is thus given by \eqref{trivialcase}.]

In the case of  
   $B-A<m_2< 2B$, we instead have:
 \begin{align}
 \rho(z)=&
 \begin{cases}
 \frac{1}{2 \lambda} \hspace*{6mm} z \in [-A,  m_2-B]
 \\ \nonumber
 \frac{1}{3 \lambda} \hspace*{6mm} z \in [ m_2-B,B]
 \end{cases}
. \end{align}%
 whereas in the case
 $0<m_2<B-A$, we find:
 \begin{align}
 \rho(z)=&
 \begin{cases}
 \frac{1}{3 \lambda} \hspace*{6mm} z \in [-A, m_2+A]
 \\ \nonumber
 \frac{1}{2 \lambda} \hspace*{6mm} z \in [ m_2+A,B]
 \end{cases}
 ,\end{align}

\begin{figure}
\begin{center}
 \includegraphics[width=0.9\textwidth] {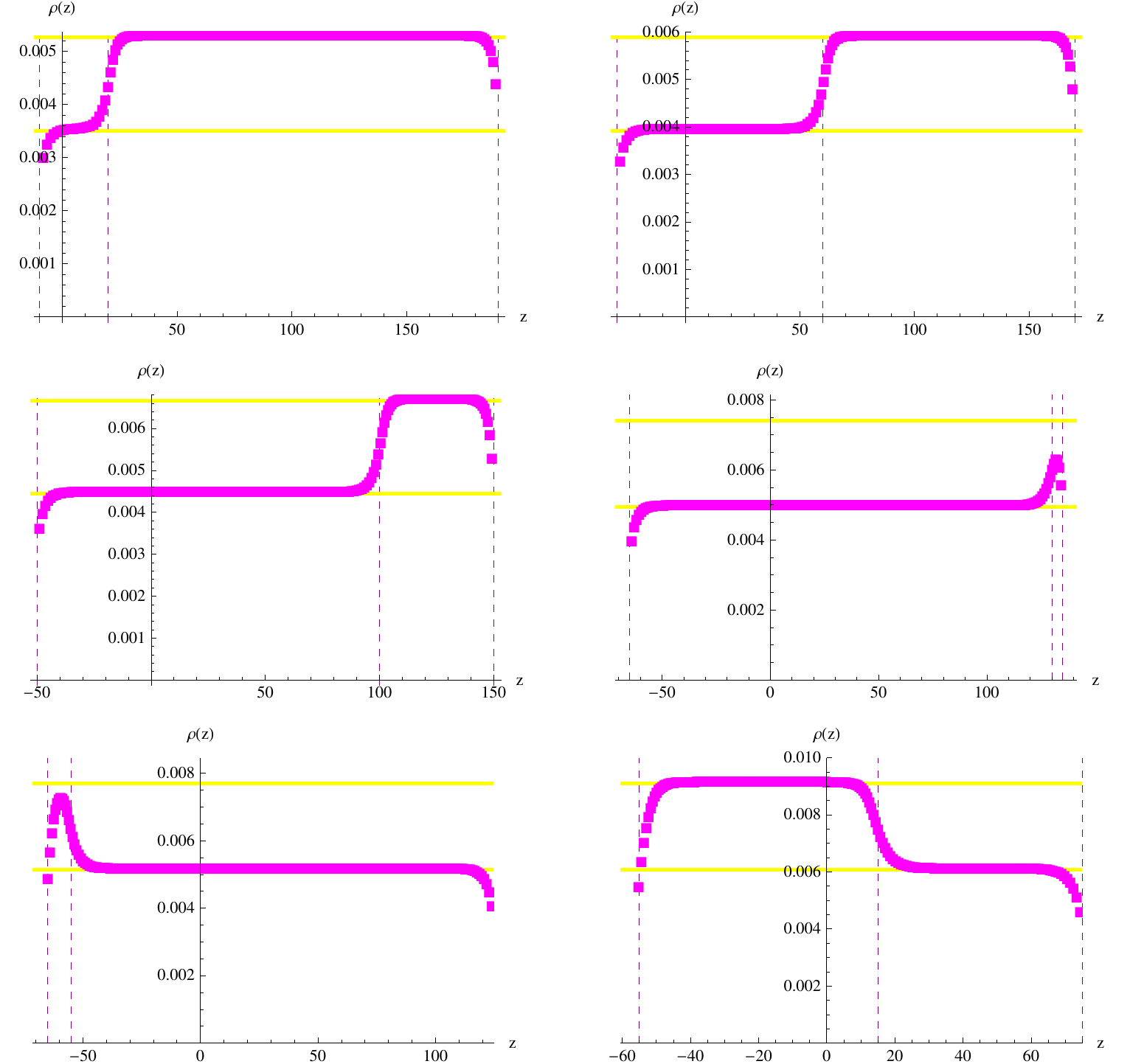}
\caption{\small Numerical solutions for large $m_1$ with $m_2 >0$. Here, $\zeta \in \{ 0.9, 0.7, 0.5, 0.35, 0.3, 0.1\} \frac{m}{2}$.}
\label{fig:mtilde_int_special_case1}
\end{center}
\end{figure}

 Normalisation together with the integral equations allows us to fix both interval endpoints in terms of $m_2,\lambda$ and the complete expression for the eigenvalue density in the case where $m_1>2A$ is then given by:

\begin{align}
\label{bqq}
\rho(z)=& \begin{cases}
 \frac{1}{2 \lambda} \hspace*{6mm} z \in [-\lambda, \hspace*{10mm} 2m_2-3\lambda]
 \\
 \frac{1}{3 \lambda} \hspace*{6mm} z \in [ 2m_2-3\lambda ,\hspace*{2mm}3\lambda-m_2]
 \end{cases} 
 \hspace*{6mm}  \frac{m_2}{2}<\lambda<m_2
 \end{align}
and
 \begin{align}
\label{aqq}
 \rho(z)=&
 \begin{cases}
 \frac{1}{3 \lambda} \hspace*{6mm} z \in [-m_2  , \hspace*{2mm} 2m_2]
 \\ 
 \frac{1}{2 \lambda} \hspace*{6mm} z \in [ 2m_2,\hspace*{4mm}2\lambda]
 \end{cases} \hspace*{15mm} \lambda>m_2\ .
\end{align} 
As seen in figure \ref{fig:mtilde_int_special_case1}, both of these cases agree well with numerics, and there is a phase transition at the point $m_2=\lambda$, as  expected.

Having determined $A$ and $B$, we can now check the region of validity of the solution.
For the solution \eqref{bqq}, the condition $m_1>2A$ gives the additional constraint $\lambda<m_1/2$.
For the solution \eqref{aqq}, $m_1>2A$  requires $2\zeta>m/3$.
In the complete phase diagram shown in fig. \ref{fig:phase_diagram}, the
 uniform eigenvalue density  \eqref{trivialcase} is the density in  the shaded region below the lowest  blue line.
The solution \eqref{bqq}  represents the eigenvalue density  in  the triangular region above this blue line, having the green ($\lambda=m_1/2$) and black ($\lambda=m_2$) lines as the other sides.
Finally, the solution \eqref{aqq} is the eigenvalue density   in  the region above this black line,
limited by the purple lines $2\zeta=m/3$ and $2\zeta=m$.


\newpage

\subsection{General resonance structure}
In the general case, we will have resonance points originating from both interval endpoints,  
which may conveniently be expressed in the tables below. Table \ref{tab:resonancesA_general} contains all resonances originating from the leftmost interval endpoint, $-A$, whereas table \ref{tab:resonancesB_general} contains the resonances from the other endpoint, $B$.

\begin{table}[h!]
  \begin{center}
    \begin{tabular}{| l |l | l  |}
      \hline
    $z$ & $-z-m_1$ &  $m_2-z$ \\
    \hline
    $-A$ 
    &  $A-m_1 $
    &  $A+m_2$   \\
        \hline
    $A+m_2$ 
    & {\scriptsize$(-A-2m)$}
    & {\scriptsize$-A$}  \\
     $A-m_1$ 
    &{\scriptsize $-A$}
    &  $-A+2m$  \\
    \hline
     $-A+2m$ 
    &  $ A -2m - m_1$
    & {\scriptsize$A-m_1$}  \\
        \hline
     $A -2m-m_1$ 
    & {\scriptsize$-A+2m$}
    &  $-A+4 m$ \\
    \hline
         $-A+4 m$ 
    &  $ A -4 m-m_1 $
    & {\scriptsize$A-2m -m_1$}  \\
    \hline
     \end{tabular}
  \end{center}
  \caption{Table of resonance points originating from the leftmost interval endpoint, $-A$, where the first row gives the ``first order'' resonances, the second row the ``second-order'' etc. Normal-sized terms denote ``new resonances'', whereas the smaller ones  coincide with previous resonance points. The small one within round brackets  lies outside the interval of support of the density. }
  \label{tab:resonancesA_general}
\end{table}

\begin{table}[h!]
  \begin{center}
    \begin{tabular}{| l |l | l  |}
      \hline
    $z$ & $-z-m_1$ &  $m_2-z$ \\
    \hline
    $B$ 
    &{\scriptsize$(-B-m_1)$}
    & $m_2-B$  \\
        \hline
    $m_2-B$ 
    & $B-2m$
    & {\scriptsize$B$}  \\
    \hline
     $B-2m$ 
    & {\scriptsize$ m_2-B$}
    & $-B+2m +m_2$  \\
        \hline
     $-B+2m +m_2$ 
    & $B-4m$
    & {\scriptsize$B-2m$}  \\
    \hline
         $B-4 m$ 
    & {\scriptsize$ -B+2m+m_2$}
    &   $-B+4m+m_2$  \\
    \hline
     \end{tabular}
  \end{center}
  \caption{Table of resonance points originating from the rightmost interval endpoint, $B$
(same conventions as table \ref{tab:resonancesA_general}). }
  \label{tab:resonancesB_general}
\end{table}

In total, the resonance points may be written as:

\begin{align}
\label{eq:resonances_general}
\nonumber
& \begin{cases}
a_{2k_a-1}&= A-m_1-2m(k_a-1) \\
a_{2k_a}& =  -A+2k_a m
\end{cases}
\\ 
& \begin{cases}
b_{2k_b-1}& =m_2-B+2m (k_b-1)\\
b_{2k_b} & = B-2k_bm
\end{cases}
\\ \nonumber
\tilde{a}=& \quad m_2+A
,\end{align}
where $k_a, \ k_b \in \{1,2,3, \dots \}$. From these expressions for the resonance points, it is clear that $\tilde{a}$ and $b_1$ cannot both lie in the interior of the interval at the same time. Rather, the condition that $\tilde{a}$ lies in $[-A,B]$ is equivalent to $m_2 < B-A$, whereas the condition that $b_1$ lies inside the interval is equivalent to $m_2 > B-A$ (i.e. exactly the condition separating the two non-trivial phases in the simple example considered in section \ref{sec:special_case:n-0}), and we again have two separate cases to consider.
We discuss these cases in the order of increasing difficulty; thus starting with the first one.

\subsection{$m_2 < B-A$ \label{sec:mtilde_leq_B-A}}

In this case there are no resonances from the rightmost endpoint $B$. The only resonance points are then  given by the $a_{2k_a},\ a_{2k_a-1}$, together with $\tilde{a}$.

Here it is convenient to  first use the integral equation  \eqref{eq:saddlepoints_cont_rescaled} to determine $B$. This calculation is straightforward, since the argument of the sign-functions are strictly positive in the integration regime. Normalisation then forces all integrals to unity, and one finds:
\begin{align}
B=&
   \lambda \overset{x<B, \; \rightarrow =1 \text{ by norm.}}{\overbrace{\int_{-A}^{B} dx'\,\rho  (x ') \; \sign(B -x') }}
   +
      \frac{\lambda}{2}  \overset{B>-y, m_1>0, \; \rightarrow =1 \text{ by norm.}}{\overbrace{   \int_{-A}^{B} d y\,\rho  (y ) \; \sign(B+y+m_1) }}
\\ \nonumber& 
\qquad 
+ \overset{m_2<B-A, \; \rightarrow 1 \text{ by norm.}}{\overbrace{ \frac{\lambda}{2}   \int_{-A}^{B} d y\,\rho  (y )\; \sign(B+y-m_2)}}
\qquad  \qquad \qquad \qquad
=  2\lambda
,\end{align}
in the case where $m_2 < B-A$.

The interval $[-A,B]$ will as always be divided into parts by the interior resonance points, which in this case are given by the  points $\tilde{a},a_{2k_a},\ a_{2k_a-1}$, where $k_a$ is limited by the condition that $a_{2k_a-1}$ lies within the interval. This implies $ k_a \leq \frac{A-\zeta}{m}+\frac{1}{2}.$
  Let the highest integer which fulfils this be denoted by $n$, such that 
\begin{align}
  \label{eq:small-mtilde:n-def}
   n= \left[  \frac{A-\zeta}{m}+\frac{1}{2} \right]
  . \end{align}
   This means that $n$ will be the integer number of times $2 m$ fits in $[-A,\tilde{a}]$, that is, in $2(A-\zeta)+m$. (The number of even resonances will hence be equal to $n$.)  Define also $\Delta$ to be given by: 
\begin{align}
\label{eq:Delta-def}
\Delta=2(A-\zeta)+m-2nm
.\end{align}

\begin{figure}[t]
\begin{center}
 \centerline{\includegraphics[width=12cm]{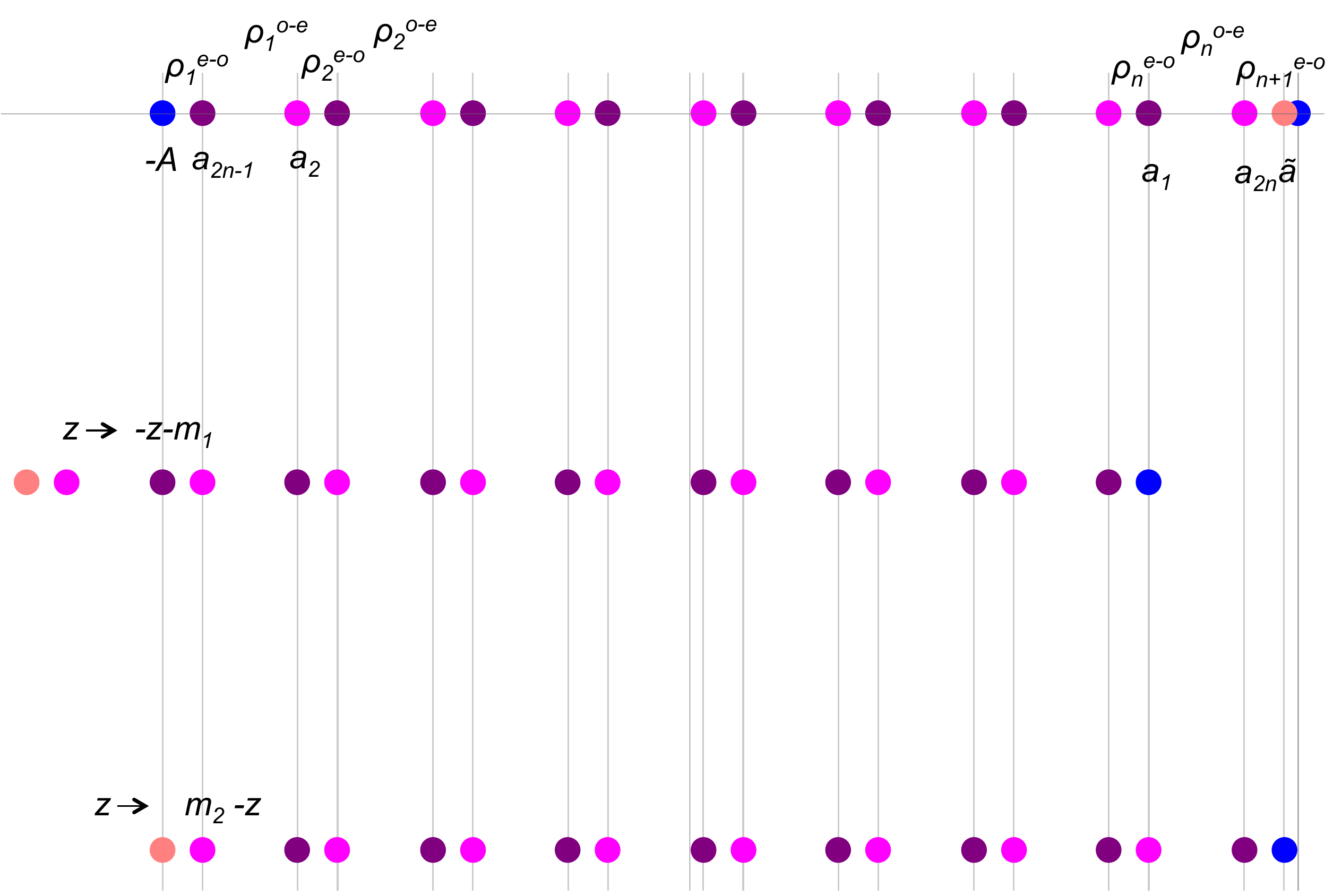}}
\caption{\label{fig:SmallFI_int_res_from_A}\small  Resonance points originating from the interval endpoint $-A$ (shown in blue), where  pink represents $\tilde{a}$, and the  purple points are the odd resonance points whereas the  magenta ones are the even ones. Below these are the same points under the maps  $z \rightarrow -z-m_1$ and $z \rightarrow m_2-z$ respectively. The density is as such divided into a piecewise constant density, made up out of two sets: the density on the regions going from odd-even resonance points ($\rho^{o-e}_k$), and the density on the regions going from the even to odd ones, denoted by $\rho^{e-o}_k$.}
\end{center}
\end{figure}

The total of $2n+1$ resonances dividing the interval will be ordered as 
$\tilde{a}>a_{2n}>a_1>a_{2(n-1)} > a_3 >\dots >a_{2n-1}>-A$. 
Just as in the cases previously considered, the eigenvalue density will be piecewise constant, and we may define different constants on the different patches of the interval, such that:
\begin{align}
\rho(z)=
\begin{cases}
\rho^{e-o}_k & \text{in the regions between an even resonance point and an odd one} \\
\rho^{o-e}_k & \text{in the regions between an odd resonance point and an even one}  \\
\frac{1}{2 \lambda} & \text{between } \tilde{a} \text{ and } B
\end{cases}
,\end{align}
as illustrated in figure \ref{fig:SmallFI_int_res_from_A}. In total, there will be $n$ $\rho^{o-e}$ and $n+1$ $\rho^{e-o}$'s. Furthermore, the saddle-point equation 
\eqref{eq:saddles_ana1_equal}  
for these two different sets of constants decouple, and we are left with:
\begin{align}
\label{eq:SDPT_mtilde_leq_B-A}
2 \rho^{e-o}_k+\rho^{e-o}_{n+1 - k} + \rho^{e-o}_{n+2-k} 
\;= & \;
\frac{1}{ \lambda} 
& \forall   k \in[1, n+1]
\\ \nonumber
2\rho^{o-e}_k  + \rho^{o-e}_{n - k} + \rho^{o-e}_{n+1-k} 
\; = & \; \frac{1}{ \lambda} 
& \forall  k \in [1, n]
,\end{align}
where the boundary conditions are given by $\rho^{e-o}_0=\rho^{o-e}_0=0$.

These may be solved by a polynomial Ansatz in $k$, where one finds the general solution to be a linear function in $k$. The constant terms are then forced to vanish by the boundary conditions, giving the two sets of linear terms as:
\begin{align}
\rho_k^{e-o}= 
 \frac{k }{\lambda (3  +2  n) }
\hspace*{3cm}
\rho_k^{o-e}=  \frac{k }{\lambda(1 +2  n)}
.\end{align}

The normalisation condition on the eigenvalue density may be written in terms  of $\rho_k^{e-o}, \rho_k^{o-e}$, such that:
\begin{align}
\label{eq:norm_small_FI_1}
1=\Delta \sum_{k=1}^{n+1} \rho_k^{e-o} +(2m- \Delta) \sum_{k=1}^{n} \rho_k^{o-e} -\frac{\tilde{a}}{2 \lambda}+1
.\end{align}
Together with the definition of $\Delta$, this allows us to solve for $A$ and $\Delta$ in terms of $m,\ \zeta$ and $n$, and we find:
\begin{align}
A=& \; m( 2  n+1) -2 \zeta  (2 n (n+2)+1)\ ,
\\ \nonumber
\Delta =& \;   (2 n+3) (m-2 \zeta  (1+2  n))
,\end{align}
with $B=2\lambda $.

Having  this expression for $A$, we may find the critical point at which the value of $n$ shifts. At the very site where $n= \frac{A-\zeta}{m}-\frac{1}{2}$, that is, we enter the region with a specific $n$, we find the following condition on $2\zeta$:
\begin{align}
\label{eq:phase_A_nlimit}
2\zeta \;=\; \frac{m}{2n+1}
.\end{align}
However, it is interesting to notice that this is independent of the coupling, and so transitions between different phases of this kind, with only resonance points originating from the leftmost interval endpoint, $-A$, will only occur with shifts in  $\zeta/m$.

Next, consider phase transitions into a phase where the resonance points from $B$ starts to move inside the interval.
For fixed values of $m,\zeta$ (and such also $n$), the condition that these resonances will remain on the outside of the interval corresponds to $m-2\zeta < B-A$, which in terms of the coupling may be written as:
\begin{align}
\label{eq:coupling_limit}
(n+1) (m-2 \zeta  (n+1))  < \lambda
.\end{align}
Therefore, as $\lambda$ decreases, we  leave this regime and enter the next one, where resonances appear from both endpoints.

The general eigenvalue density in this case, for an FI-parameter in $ \zeta < \frac{m}{2}$ and coupling satisfying equation \eqref{eq:coupling_limit}, will be given by:
\begin{align}
\rho(z) =
\begin{cases}
 \frac{k }{\lambda (3  +2  n) }  &\forall z \in [a_{2(k-1)}, a_{2(n-k)+1}] 
 \hspace*{1.9cm} k \in [1 \;, \;n+1]\\
  \frac{k }{\lambda(1 +2  n)} &\forall z \in [a_{2(n-k)+1},a_{2k}]
  \hspace*{2.55cm} k \in [1 \; , \; n] 
\label{cuarenta}\\
\frac{1}{2 \lambda}  &\forall z \in [\tilde{a},2 \lambda]
\end{cases}
,\end{align}
where $a_0$ and $ a_{-1}$ should be interpreted as $-A$ and $\tilde{a}$ respectively,  and  $n$  is defined by equation \eqref{eq:small-mtilde:n-def}.

\subsection{$m_2> B-A$ \label{sec:mtilde_geq_B-A}}

This situation is more complicated. Although $\tilde{a}$ now lies outside the interval, there are now interior resonance points originating from the rightmost interval endpoint $B$ as well. 
The complete set of resonances in this case is given by the expressions in equation \eqref{eq:resonances_general}, namely: 

\begin{align}
 \begin{cases}
a_{2k_a-1} \quad \;= A-m_1-2m(k_a-1) 
\qquad  &
\forall  \; 0 < k_{a} \;<\;  \frac{A-\zeta}{m} +\frac{1}{2} \\
a_{2k_a} \qquad =  -A+2k_a m
\qquad & 
\forall  \; 0 <  k_{a} \; <\frac{A+B}{2m}
\end{cases}
\\ \nonumber
 \begin{cases}
b_{2k_b-1} \; =m_2-B+2m (k_b-1)
\qquad & 
\forall  \; 0 < k_b \;<  \frac{B+\zeta}{m}+\frac{1}{2}
\\
b_{2k_b} \quad = B-2k_bm
\qquad   & 
\forall \; 0 < k_b \; < \frac{A+B}{2m}
\end{cases}
\end{align}
where $k_a,k_b \in {1, 2, \dots}$ and the upper limits comes from requiring the resonances to lie inside the interval 
of support of the eigenvalue density $[-A,B]$.

When $m_2>B-A$, it is easy  to see that 
\begin{align}
  \frac{A+B}{2m} 
\quad< \quad 
 \frac{A-\zeta}{m}+\frac{1}{2} 
\quad< \quad 
\frac{B+\zeta}{m}+\frac{1}{2}
,\end{align}
and both $k_a,k_b$ are then limited by $\frac{A+B}{2m}$. We again define $n$ as the integer part of this number,
\begin{align}
n=\left[ \frac{A+B}{2m} \right]
,\end{align}
such that $a_{2n},b_{2n}$ denotes the final even resonances. 
However, in some cases, the following odd resonances may lie within the interval $[-A,B]$ as well. This gives rise to three different
cases that are examined in appendix \ref{app} in detail.
These are:

\begin{itemize}
\item {\bf Case I ($n =\left[ \frac{A+B}{2m} \right], \quad  \frac{A-\zeta}{m}-\frac{1}{2} <  \frac{B+\zeta}{m}-\frac{1}{2} <  n $):}
\begin{align}
\label{casoI}
A=&  \lambda\, \frac{2  n+1}{n+1}-2 \zeta  n
\\ \nonumber
B =&  \lambda\, \frac{2  n+1}{n+1}+2 \zeta  n
\end{align}

\item {\bf Case II ($n =\left[ \frac{A+B}{2m} \right], \quad \frac{A-\zeta}{m}-\frac{1}{2} < n <  \frac{B+\zeta}{m}-\frac{1}{2}$):}
\begin{align}
\label{casoII}
A=&  \lambda\, \frac{2  n+1}{n+1} -2 \zeta  n \\ \nonumber
B=& - m + 2\zeta (n+1) + \lambda \, \frac{  2 n+3}{n+1}  
\end{align}

\item {\bf Case III ( $n =\left[ \frac{A+B}{2m} \right], \quad  n < \frac{A-\zeta}{m}-\frac{1}{2} < \frac{B+\zeta}{m}-\frac{1}{2}$):}
\begin{align}
\label{casoIII}
A =& -m-2 \zeta  (n+1)+\lambda\  \frac{ 2 n+3}{n+1}
\\ \nonumber
B =& -m+2 \zeta  (n+1)+\lambda\ \frac{  2 n+3}{n+1}
\end{align}
\end{itemize}

\subsection{Phase transitions \label{faso}}

The condition $m_2 > B-A$ also manifests differently in the three different cases. In cases {\bf I} and {\bf III}, this turns out to be completely independent of the coupling, whereas this is not the situation in case {\bf II}. However, this is to be expected since the second situation is quite similar to the situation with resonances only from the interval endpoint $-A$, considered in section \ref{sec:mtilde_leq_B-A}. The condition found upon the coupling in case {\bf II} is precisely 
\begin{align}
\label{eq:coupling_cond_caseII}
\lambda \leq \lambda^A_c\ ,\qquad \lambda^A_c\equiv (n+1)(m-2\zeta(n+1))\ 
,\end{align}
which is precisely opposite to the condition \eqref{eq:coupling_limit} in section \ref{sec:mtilde_leq_B-A}. Thus, as the coupling decreases for a fixed $m,\zeta$ and, therefore, $n$, eventually,  the inequality is saturated in \eqref{eq:coupling_cond_caseII}, the resonances from $B$ move inside the interval and a phase transition occurs, leading to  case {\bf II} of this section. 

The conditions for the two other cases considered in this section are 
\begin{align}
\text{Case {\bf I}:} &\qquad \zeta< \frac{m}{2(n+1)}
\\ \nonumber
\text{Case {\bf III}:} & \qquad \zeta < \frac{m}{2(2n+3)}
\end{align}

In this situation, where $\zeta < \frac{m}{2}$, and $m_2>B-A$, there are clearly phase transitions. This is   expected, since they do appear in the situation with vanishing FI-parameter. 
However, there are  different kinds of phase transitions: both in between the three different cases (described in detail in sections \ref{sec:case_1} -- \ref{sec:case_3}), within the same value of $n$, and one where the value of $n$ changes. 

Let us start with some fixed value of $n$, such as $n=\left[ \frac{A+B}{2m}\right]$, and let us consider the case where $n$ is larger than $\frac{B+\zeta}{m}-\frac{1}{2}$.
Then, as the coupling increases, so does $\frac{B+\zeta}{m}-\frac{1}{2}$ (growing linearly with coupling) for some fixed $n$. At some point this will surpass this $n$, and a phase transition occurs that will take us to case {\bf II} above. This will happen for the coupling:
\begin{align}
\label{eq:crit_I_II}
\lambda^c_{{\bf I}\rightarrow {\bf II}} =\frac{1}{2} (n+1) (m-2 \zeta )
,\end{align}
which in the case of vanishing FI-parameter simply corresponds to the situation where another factor of $m$ fits inside the interval. 

From there onwards, as the coupling grows further, so will the factor $\frac{A-\zeta}{m}-\frac{1}{2}$ (also growing linearly with $\lambda$), and at the point 
\begin{align}
\label{eq:crit_II_III}
\lambda^c_{{\bf II}\rightarrow {\bf III}} =\frac{1}{2} (n+1) (m+2 \zeta )
,\end{align}
another phase transition occurs, taking us into case {\bf III}.
It is worth noticing that this phase transition does not occur in the case of vanishing FI-parameter, but rather coincides with the one between case I and II, simply because the second case, where $n \in [\frac{A-\zeta}{m}-\frac{1}{2}, \frac{B+\zeta}{m}-\frac{1}{2}]$, never occurs for $\zeta=0$, since this interval then is empty. 

The final phase transition then occurs as one goes from case {\bf III} back to case {\bf I}, $\frac{A+B}{2m}$ has grown to the point that the integer part of it changes, and that  $n$  increases with one. This happens at the point $\frac{A+B}{2m}=n+1$, occurring at
\begin{align}
\label{eq:crit_III_I}
\lambda^c_{{\bf III} \rightarrow {\bf I}} =& \frac{m (n+1) (n+2)}{2 n+3}
.\end{align}
This corresponds to the result obtained in \cite{Anderson:2014hxa} for the transition from $m$ fitting a total of $2n+1$ times inside the interval $[-A,A]$ to $m$ fitting a total of $2n+2$ times, for $\zeta=0$. (Again, in this special case, this corresponds to another multiple of $m$ fitting inside the interval.)

In order to determine the order of these phase transitions, we need to 
study  the analytic properties of the free energy, 
\begin{align}
F=-\frac{1}{RN_1N_2} \ln Z
,\end{align}
 at the critical values of the coupling.
It turns out that both the first- and second-order derivative of $F$  \emph{are} continuous, whereas \emph{the third-order derivative is not. } Let us denote by $\Delta F$ the difference between the subcritical and supercritical free energy in each phase transition.
For the phase transitions occurring between the case with only resonances from $-A$, for some given $n$, (presented in section \ref{sec:mtilde_leq_B-A}) to  case {\bf II} with this same $n$ presented above, occurring at $\lambda_c^{A }=(n+1)(m-2\zeta(n+1))$, we find:
\begin{align}
\partial_\lambda \; \Delta F \Big|_{\lambda_c^{A}}  =& \; \partial^2_\lambda \; \Delta F \Big|_{\lambda_c^{A}}  = \;0\ ,
\\ \nonumber
\partial^3_\lambda\;  \Delta F \Big|_{\lambda_c^{A}}  =&\; \frac{2}{(n+1)^4 (m-2 \zeta  (n+1))^2}
.\end{align}
Note that the apparent singularity at $m=2 \zeta  (n+1)$ is outside the region where these solutions apply,
$m/(2n+3)<2\zeta < m/(2n+1)$.

Similarly, at the critical points between the cases {\bf I}, {\bf II} and {\bf III} above (described by equations \eqref{eq:crit_I_II} -- \eqref{eq:crit_III_I}), we once again find a discontinuity at the third derivative, whereas all lower derivatives are continuous:

\begin{align}
\partial^3_\lambda\;  \Delta F \; \Big|_{\lambda_c^{ {\bf I}(n) \rightarrow {\bf II}(n) }} =&
\frac{32}{(n+1)^3 (m-2 \zeta )^2}
\\ \nonumber 
\partial^3_\lambda\;  \Delta F  \; \Big|_{\lambda_c^{{\bf II}(n) \rightarrow {\bf III} (n)}} =&
\frac{32}{(n+1)^3 (m+2 \zeta )^2}
\\ \nonumber 
\partial^3_\lambda\;  \Delta F  \; \Big|_{\lambda_c^{{\bf III}(n) \rightarrow {\bf I}(n+1)}} =&
-\frac{2 (2 n+3)^5}{m^2 (n+1)^4 (n+2)^4}
.\end{align}

\noindent Therefore, just as in the case with vanishing FI-parameter, these phase transitions are all of third order. 

\medskip

Finally, one may also look at transitions between solutions \eqref{cuarenta}
of section  \ref{sec:mtilde_leq_B-A}  with different
$n$, as the FI parameter $\zeta $ is increased along lines of constant $\lambda $.
The solution \eqref{aqq} represents the case $n=0$ and is valid in the region $2\zeta>m/3$, $\lambda>m_2$. The case $n=1$ can be readily found from the general formulas of section  \ref{sec:mtilde_leq_B-A}. We obtain
  \begin{align}
\label{aqqq}
 \rho(z)=&
 \begin{cases}
 \frac{1}{5 \lambda} \hspace*{6mm} z \in [-A , \ a_1]
 \\ 
 \frac{1}{3 \lambda} \hspace*{6mm} z \in [ a_1,\ a_2]
\\ 
 \frac{2}{5 \lambda} \hspace*{6mm} z \in [ a_2, \ m_2+A]
\\ 
 \frac{1}{2 \lambda} \hspace*{6mm} z \in [ m_2+A, \ 2\lambda]
 \end{cases} \hspace*{15mm} \lambda> 2m-8\zeta\ ,\quad \frac {m}{5}<2\zeta<\frac{m}{3}\ ,
\end{align} 
with
\begin{align}
A=3m-14\zeta\  ,\quad a_1=2m-16\zeta\ , \quad a_2=14\zeta-m\ .
\end{align}
In crossing the line $2\zeta =m/3$ from lower to higher values of $\zeta $, the solution changes from \eqref{aqqq} to \eqref{aqq} (in fig. 
\ref{fig:phase_diagram}, this corresponds to crossing the purple line at $2\zeta =m/3$ that begins at an hexa-critical point). The discontinuity in the free energy shows up in the third derivative
with respect to $\zeta $. Again, for convenience we first compute  $\partial_\lambda F$
which has a simple local expression, given by $-\langle z^2\rangle/\lambda^2$.
We find that first and second order $\zeta $ derivatives are continuous, whereas
\begin{align}
\partial^3_\zeta \left(\Delta \partial_\lambda F\right)\bigg|_{2\zeta=\frac{m}{3}} = \frac{432}{\lambda^3}\ .
\end{align}
More generally, for the transition from the $n-1\to n$ solutions \eqref{cuarenta} we find
\begin{align}
\partial^3_\zeta \left(\Delta \partial_\lambda F\right)\bigg|_{2\zeta=\frac{m}{2n+1}} = \frac{16(2n+1)^3}{\lambda^3}\ .
\end{align}
Thus the  quantum phase transitions between these  phases are also of the third order.

\section{Summary of the Analytically Continued Model  \label{sec:Conclusion}}

\subsection{ Case $\lambda_1=\lambda_2$}

We have herein seen that the phase structure found in \cite{Anderson:2014hxa} is significantly enriched when the theory is deformed by a FI-parameter, $\zeta$. The saddle-point equations then loose the reflection symmetry  present at $\zeta = 0$, and the behavior in the decompactification limit is highly dependent on the value of this new parameter. 
As $\zeta \rightarrow 0$, we recover the results of \cite{Anderson:2014hxa}.

For an FI-parameter large enough, $\zeta \geq \frac{m}{2}$, the theory is free from phase transitions.
However, for $ |\zeta| <  \frac{m}{2}$, phase transitions appear. 
The precise appearance of the eigenvalue density and the position of  the phase transitions in phase space depend on $\lambda$ and $\zeta$ in relation to $m$.

\begin{figure}[t]
\begin{center}
 \centerline{\includegraphics[width=11cm]{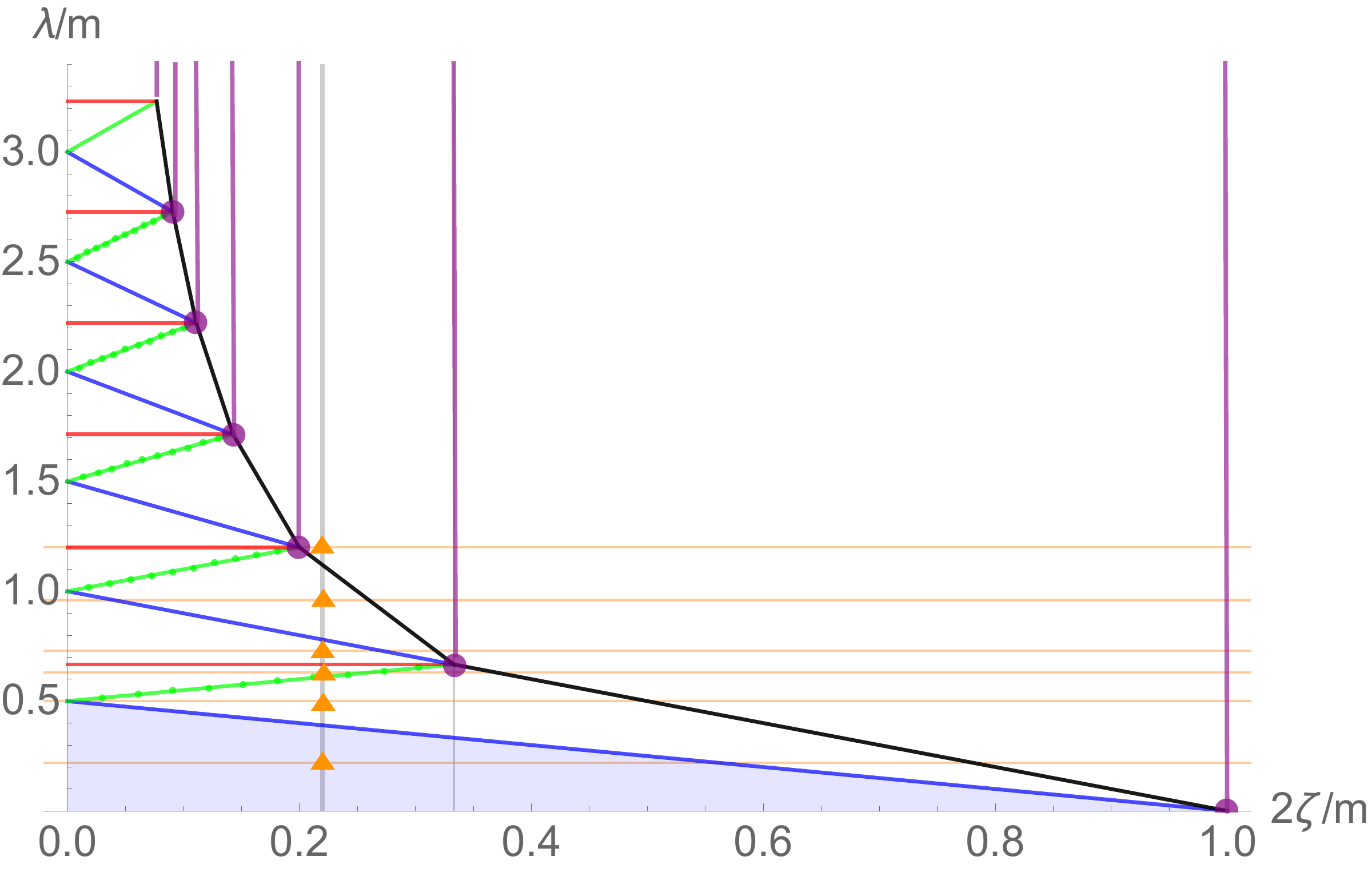}}
\caption{\label{fig:phase_diagram}\small  Phase diagram of the analytically continued model. Black critical lines separate phases  of case {\bf II} (below) from phases with only resonances from $-A$ (above). The shaded region corresponds to the constant eigenvalue density   \eqref{trivialcase}.
Crossing  the blue,  green  and  red  lines corresponds to phase transitions from cases with resonances from both endpoints, ${\bf I} \rightarrow {\bf II}$, ${\bf II} \rightarrow {\bf III}$ and ${\bf III} \rightarrow {\bf I}$ (where $n$ increases by one), described in section \ref{sec:mtilde_geq_B-A}.
The purple dots are hexa-critical points.
The purple (vertical) lines separate phases described in section  \ref{sec:mtilde_leq_B-A}  differing in one unit in the value of $n$, representing solutions with  $2n+1$ resonances  from the leftmost interval endpoint.
At the orange triangles  the eigenvalue density has been
computed numerically and compared to theory (see  fig. \ref{fig:phases_ex}). }
\end{center}
\end{figure}

To illustrate this dependence, one may consider a phase diagram with the dimensionless axes $\lambda/m$ and $2\zeta/m$
(figure \ref{fig:phase_diagram}). Phase transitions then occur on certain critical lines in this phase space. 
There are no phase transitions in the region where the FI-parameter satisfies $\zeta \geq \frac{m}{2}$, and so there is no use to show the phase diagram further than $2\zeta/m =1$.
For $2\zeta/m<1$, the only phase with no resonances is the constant eigenvalue density  \eqref{trivialcase},
occurring in the shaded region of fig. \ref{fig:phase_diagram}. 
At the purple dots, there are six coexisting phases: they are hexa-critical points.

\begin{figure}[t]
\begin{center}
 \centerline{\includegraphics[width=8cm]{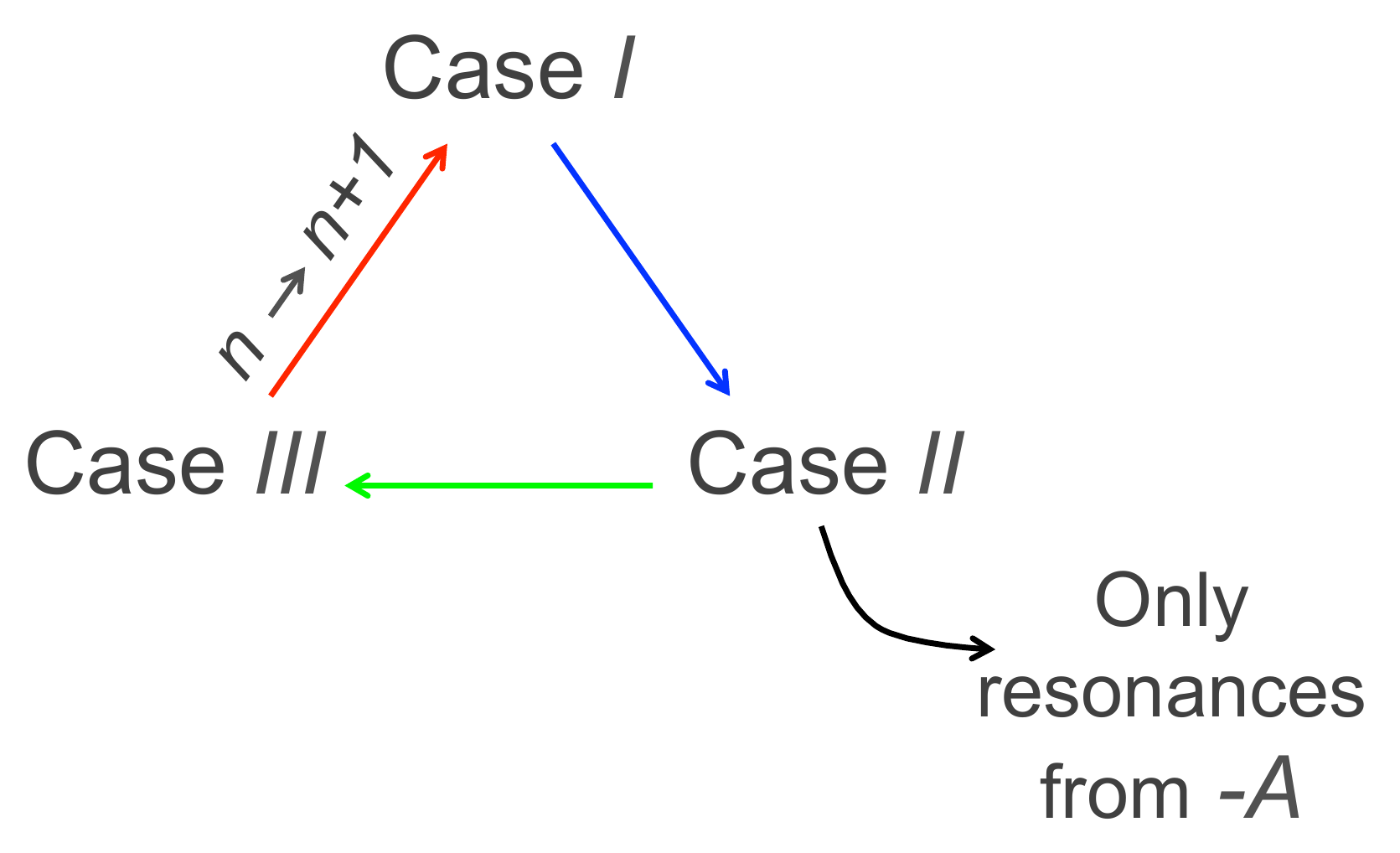}}
\caption{\label{fig:phase_summary}\small  Sequences of phase transitions  in the theory at constant $\zeta $. As $\lambda $ grows, the cycle continues until the value of $n^{\text{max}}$ where, for given $m, \zeta$,  one has $\lambda_c^{{\bf II} \rightarrow A}(n^{\text{max}}) <\lambda_c^{{\bf II} \rightarrow {\bf III}}(n^{\text{max}})$, after which the cycle ends.  }
\end{center}
\end{figure}

It is worth noting that the phase transitions between cases {\bf I},  {\bf II} and {\bf III} only occur in a specific order (see fig.  \ref{fig:phase_summary}). The number of phase transitions undergoing as the coupling $\lambda$ runs from $0$ to infinity depends on the value of $2\zeta/m$, and the maximal  $n^{\rm max}$ that will occur
is given by the largest $n$ fulfilling $\lambda^c_{{\bf III} \rightarrow {\rm {\bf I}}} (n) < \lambda_c^A(n)$, which, by using equations \eqref{eq:coupling_cond_caseII} and \eqref{eq:crit_III_I}, implies:
  \begin{align}
  \label{eq:nmax}
 n^{\rm max}=\left[ \frac{m}{4\zeta}-\frac{1}{2} \right]
 . \end{align}
 The number of phases is then given by $3(n^{\rm max}+1)$,
It approaches infinity as $2\zeta/m \rightarrow 0$. This is to be expected in order to match the $\zeta =0$ case \cite{Anderson:2014hxa}. However, for any   non-vanishing FI-parameter, there is only  a finite number of phase transitions in the theory (and no phase transition for $2\zeta\geq m$).

 The grey, vertical line in figure \ref{fig:phase_diagram} at $2\zeta/m=0.22$ illustrates an example on the phase structure of the model, and how to read the diagram. 
 For $2\zeta/m=0.22$, one has $n^{\rm max}=1$ by \eqref{eq:nmax}, hence $6$ different phases
along the grey line.
  Starting at the bottom of the diagram, we are in the trivial phase (case {\bf I} with $n=0$).   As the coupling grows we moves upwards along the grey line until we cross the first blue line. This corresponds to a phase transition into case {\bf II} with $n=0$. After crossing the green line, we enter {\bf III} with still $n=0$, and when crossing the red line, we move back into case {\bf I}, but now with $n=1$. Crossing the next blue line  takes us to case {\bf II}, $n=1$.
 However, instead of crossing another green (and thereafter red) line, we rather cross a black line.
Beyond this point, all resonances from the rightmost interval endpoint  move outside the interval of support of the eigenvalue density, and we thus get into the phase described in section \ref{sec:mtilde_leq_B-A} with $n=1$, having resonances only from the leftmost interval endpoint $-A$. 

\begin{figure}[t]
\begin{center}
 \centerline{\includegraphics[width=11cm]{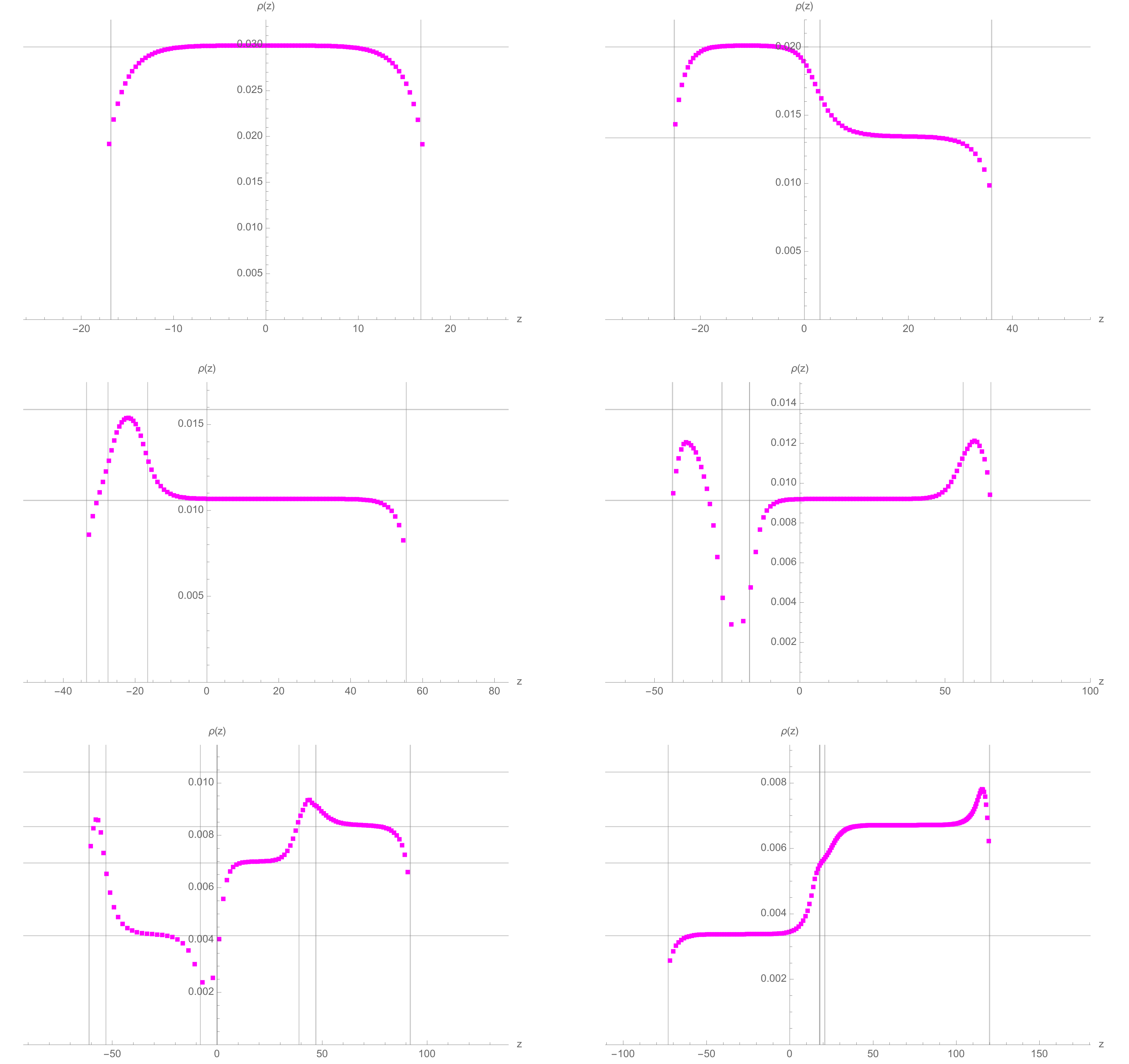}}
\caption{\label{fig:phases_ex}\small  Eigenvalue densities determined numerically for the six phases present with $2\zeta/m=0.22$ and $\lambda/m \in \{0.22, 0.5,0.62,0.73, 0.96, 1.2  \}$, compared  with  eigenvalue density determined analytically, for the cases ${\bf I}_{n=0}, {\bf II}_{n=0}, {\bf III}_{n=0}, {\bf I}_{n=1}, {\bf II}_{n=1}, A_{n=1}$ (grey lines). [The smooth edges of the eigenvalue density are due to the finite radius $R$ used in the calculations. Here, $mR=50$. For higher
 $mR$ the numeric is unstable.] }
\end{center}
\end{figure}

   As the coupling grows further, no new resonance points  enter the interval, and the system never leaves this phase. The orange horisontal lines in figure \ref{fig:phase_diagram}, together with the orange triangles, represents points at which numerical calculations have been made in order to compare with theoretical calculations. The eigenvalue density for the six phases present for values of $m, \zeta$ such that $2\zeta/m=0.22$ are presented in figure \ref{fig:phases_ex}, and we find an excellent agreement with our theoretically derived densities in sections \ref{sec:mtilde_leq_B-A}, \ref{sec:mtilde_geq_B-A} and appendix \ref{app}. The smoothness of the curves visible in figure  \ref{fig:phases_ex} has to do with finite-size effects. 

As a check,  using equations \eqref{casoI}-\eqref{casoIII}, we may compute the values of $n$ in these different phases, and these results are summarised in Table \ref{tab:n_vals}.
The results follow the expected pattern, where $n$ changes in
going from  case {\bf III} to case {\bf I}. 

\begin{table}[h!]
  \begin{center}
    \begin{tabular}{| l |l | l  | l  | l  | l  | l  |}
      \hline
    $\lambda/m$ 
    & 0.22 
    &  0.5
    &0.63
    & 0.73
    & 0.96
    & 1.2
    \\
    \hline
    $n$ 
    &  $\left[ 0.22 \right] =0 $
    &  $\left[ 0.61 \right] =0 $
    & $\left[ 0.89 \right] =0 $
    &  $\left[ 1.095  \right] =1 $
    &  $\left[ 1.53  \right] =1 $
    &  $\left[ 1.85 \right] =1 $  \\
        \hline
        Phase
    &  {\small ${\bf I}_{n=0} $}
    &  {\small${\bf II}_{n=0} $}
    &  {\small${\bf III}_{n=0} $}
    &  {\small${\bf I}_{n=1} $}
    &  {\small${\bf II}_{n=1} $}
    &  {\small(only $A$-resonances)$_{n=1} $}  \\
        \hline
     \end{tabular}
  \end{center}
  \caption{Sequence of transitions and corresponding $ n= \left[ \frac{A+B}{2m}\right] $ in cases {\bf I} -- {\bf III}  and $n= \left[ \frac{A-\zeta}{m}-\frac{1}{2}\right] $ in the phase with only resonances from one endpoint.  }
  \label{tab:n_vals}
\end{table}


The results of section \ref{sec:mtilde_geq_B-A} (and appendix \ref{app})  explain some of the  peculiar properties of this model in the special case of $\zeta=0$:
as resonances from both the left- and right interval endpoints appear in the interior of the interval, the interval is divided into four qualitatively different regimes: two in between resonances originating from the same interval endpoint, and two in between resonances originating from different interval endpoints.
As the FI-parameter goes to zero, these regimes become pairwise indistinguishable from one another due to the presence of reflection symmetry. 
The different origins of these regimes are indeed visible in the eigenvalue density, as ``odd'' and ``even'' patches  of the eigenvalue density behave significantly different from one another; in addition,   the critical couplings at which phase transitions occur are different for ``odd'' and ``even'' phases (where mass parameter $m$ fits inside the interval an even- or an odd number of times).
 This is all explained here by the fact that these regimes  actually originate from different sets resonances; some originating from the leftmost endpoint, and some from the rightmost endpoint of the interval.
In this  way, the general solution with $\zeta\neq 0$ explains a phenomenon which appears to have no deep reason in the case of vanishing FI-parameter.

 Furthermore, one may note that the 
green and blue lines  in figure \ref{fig:phase_diagram} coincide when $\zeta=0$. Here, the only phase transitions occurring are the ones between the cases {\bf I} and {\bf III}. This is required to agree with previous results for $\zeta=0$ \cite{Anderson:2014hxa}.

\bigskip

%
%
%

\newpage


\subsection{ Case  $\lambda_1 \neq \lambda_2$ \label{diflam}}

\subsubsection{First analytic continuation \label{rtu}}

It is interesting to ask what happens in the case when the couplings are different.
We assume $\lambda_1,\ \lambda_2$ are real (as discussed, for general complex values of
$\lambda_1,\ \lambda_2$ the saddle-point equations become complicated to solve, even numerically).
Then we have to compute the partition function in the convergence region where  $\lambda_{1,2} > 0$.
We further assume, with no loss of generality, $\lambda_1 > \lambda_2 $.

First, let us consider the limit of very large FI-parameter. In the case of $\zeta \rightarrow \infty$, both shifted terms of the saddle-point equations \eqref{eq:saddles_ana1} vanish, and we are left simply with 
\begin{align}
\label{constantes}
\rho_x = \frac{1}{2\lambda_1}\ ,
\hspace*{3cm}
\hat{\rho}_y =\frac{1}{2 \lambda_2}
.\end{align}
Using the integral equations \eqref{eq:saddlepoints_cont_rescaled} with $x=B, y=D$, together with the normalisation conditions for the densities, one finds the interval endpoints to be given by:
\begin{align}
B=\,\; D= \lambda_1+\lambda_2
\\ \nonumber
A=-C=\lambda_1-\lambda_2
\end{align}
Consider what happens when the FI-parameter decreases. Then there are two points where the above solution \eqref{constantes} may cease to be valid:

\begin{itemize}
\item A resonance point moves inside the region of support for the eigenvalue density. The first point 
to do this is the one originating from the leftmost interval endpoint of \emph{the other} interval. That is, $A-m_2$ (or $C-m_2$) moves inside the interval $[-C,D]$ (or equivalently $[-A,B]$).
\item The  $\sign$-functions change inside the integration regime in \eqref{eq:saddlepoints_cont_rescaled}.
\end{itemize}

Given the solution in the case $\zeta \rightarrow \infty$,
 we find that the first one of these situations occur precisely when $m_2 \leq 0$ and, just as in the case with equal couplings, there will be a phase transition in the theory as soon as $\zeta \leq \frac{m}{2}$. 

As for the second possible point of failure for the solution \eqref{constantes} for large $\zeta$, one may easily show that the sign-functions will not change signs inside the integration regimes until $\zeta \; \leq \; \frac{m}{2}-  \lambda_2$ , $
\zeta  \; \leq  \; \frac{m}{2} - \lambda_1 
$,
which for positive couplings always  happen \emph{after} the phase transition originating from resonance points entering the intervals.
This shows that the theory with $\zeta\geq \frac{m}{2}$ does not have phase transitions, generalizing
the result found in section \ref{sec:large_FI} to the case of arbitrary $\lambda_{1},\ \lambda_2 >0$.
 

We now derive the solution in this new phase, where $\zeta$ decreases just below $\frac{m}{2}$.
Assume  $\zeta =\frac{m}{2}-u$, for some $u$, satisfying $0<u < \frac{m}{2}$ (together with some other conditions which will shortly be specified). Then one finds 
\begin{align}
m_1=& 2m-2u \; (>0)
\\ \nonumber
m_2 =& 2u
.\end{align}
(It is here worth noting that both terms shifted by $m_1$ in the saddle-point equations \eqref{eq:saddlepoints_cont_rescaled} will vanish, at least for small $u$, since we know $A=-C$ in the limiting case of $u=0$.)
 
 We wish to consider a situation where the resonance point 
 from the leftmost interval endpoint of one interval lies in the interior of the other one (i.e. $A+m_2 < D \; ,\; C+m_2 < B$), which gives us the additional conditions on $u$ mentioned above.  This is equivalent to saying that the resonance points originating from the rightmost endpoints of the intervals will lie outside the region of support for the densities. 
 As long as these conditions are satisfied, the integral equations may once more be used straight away to obtain 
\begin{align}
B=D=\lambda_1+\lambda_2
,\end{align}
just as in the case of $\zeta \geq \frac{m}{2}$.
 
The intervals $[-A,B], [-C,D]$ will then be divided into two parts each, and using a piecewise constant Ansatz for the eigenvalue density on these patches, together with the normalisation condition, one obtains:
\begin{align}
\rho_x (z) =&
\begin{cases}
\frac{1}{3\lambda_1} \qquad \forall z \in [-\lambda_1+\lambda_2-2u \; ,\;-\lambda_1+\lambda_2+4u]
\\ \nonumber
\frac{1}{2\lambda_1} \qquad \forall z \in [-\lambda_1+\lambda_2+4u \; ,\;\lambda_1+\lambda_2]
\end{cases}
\\ \nonumber
\hat{\rho}_y (z) =&
\begin{cases}
\frac{1}{3\lambda_2} \qquad \forall z \in [-\lambda_2+\lambda_1-2u \; ,\; -\lambda_2+\lambda_1+4u]
\\ \nonumber
\frac{1}{2\lambda_2} \qquad \forall z \in [-\lambda_2+\lambda_1+4u \; ,\; \lambda_1+\lambda_2]
\end{cases}
.\end{align}

Thus, as the FI-parameter decreases just below $\frac{m}{2}$, a phase transition occurs, even though $\lambda_1 \neq \lambda_2$. Additional phase transitions are undergone
as resonance points continue moving inside- or outside of the region of support of the eigenvalue densities. 
The qualitative behaviour of the theory is therefore similar to the $\lambda_1=\lambda_2$ case studied 
in previous sections.

\subsubsection{Second analytic continuation \label{segunda}}

Consider for simplicity $m_1=m_2\equiv m$, i.e. $\zeta =0$.
The saddle-point equations \eqref{eq:saddlepoints_cont2} imply the following functional equations:
\begin{align}
\label{second}
\rho_x(x) =& \frac{1}{2  \alpha_1} + \frac{1}{2} \rho_y(x+m)
+ \frac{1}{2} \rho_y(x-m)\ ,
\\
\rho_y(y) =& \frac{1}{2  \alpha_2} + \frac{1}{2} \rho_x(x+m)
+ \frac{1}{2} \rho_x(x-m)
.\end{align}
The trivial solution is
\begin{align}
& \rho_x(x) =\frac{1}{2 \alpha_1}\ ,\qquad x\in (- \alpha_1, \alpha_1) 
\nonumber \\
& \rho_y(y) =\frac{1}{2 \alpha_2}\ ,\qquad y\in (- \alpha_2, \alpha_2) 
\label{onepa}
\end{align}
and holds provided $ \alpha_1+ \alpha_2<m$.

When $ \alpha_1+ \alpha_2>m$, the solution has two patches:
\begin{align}
 \rho_x(x)=&
 \begin{cases}
  \frac{1}{2  \alpha_1}  \hspace*{11mm}\quad x \in [0,a]
 \\ \nonumber
\frac{2}{3  \alpha_1} +\frac{1}{3 \alpha_2} \hspace*{6mm} x \in [ a,A]
 \end{cases}
 ,\end{align}
\begin{align}
\label{dospa}
 \rho_y(y)=&
 \begin{cases}
\frac{1}{2  \alpha_2}
\hspace*{11mm} \quad y  \in [0, b]
 \\ 
\frac{2}{3  \alpha_2} +\frac{1}{3 \alpha_1} \hspace*{6mm} y \in [ b,B]
 \end{cases}
 ,\end{align}
\begin{align}
&a=m-B=\frac{ \alpha_1 \left(2 m- \alpha_2\right)}{2  \alpha_1+ \alpha_2}\ ,\qquad B= \frac{ \alpha_2 \left( \alpha_1+m\right)}{2  \alpha_1+ \alpha_2}\ ,
\\
&b=m-A=\frac{ \alpha_2 \left(2 m- \alpha_1\right)}{ \alpha_1+2  \alpha_2}\ ,\qquad A=\frac{ \alpha_1 \left( \alpha_2+m\right)}{ \alpha_1+2  \alpha_2}\ ,
\end{align}
where we only exhibited the region $x>0$, since in this $\zeta =0$ case the density has reflection symmetry.

Thus there is a phase transition for generic $ \alpha_1,\  \alpha_2>0$. Note that the solution requires $ \alpha_1,\  \alpha_2$ to be positive, as the eigenvalue densities must be positive.
The free energy can be computed from the formulas
\begin{align}
\partial_{ \alpha_1} F= \frac{1}{2 \alpha_1^2} \langle x^2\rangle \ ,\qquad \partial_{ \alpha_2} F= \frac{1}{2 \alpha_2^2} \langle y^2\rangle \ .
\end{align}
One can check that the resulting expressions satisfy the integrability condition $\partial_{ \alpha_1}\partial_{ \alpha_2} F=\partial_{ \alpha_2}\partial_{ \alpha_1} F$.
For the  phase with uniform density (\ref{onepa}), we find 
\begin{align}
F_{\rm i}= \frac{1}{12}( \alpha_1+ \alpha_2)\ .
\end{align}
For the phase \eqref{dospa}
\begin{align}
F_{\rm ii}=\frac{-6 \left( \alpha_1+ \alpha_2\right) m^2-3  \alpha_1  \alpha_2 m+ \alpha_1  \alpha_2 \left( \alpha_1+ \alpha_2\right)+2 m^3}{12
   \left(2  \alpha_1+ \alpha_2\right) \left( \alpha_1+2  \alpha_2\right)}\ .
\end{align}

It follows that the first and second derivative are continuous at the transition point $ \alpha_1+ \alpha_2=m$,
whereas the third derivative has a jump: 
\begin{align}
\partial_{ \alpha_1}^3 \Delta F\Big|_{ \alpha_2=m- \alpha_1} = \frac{1}{(2m- \alpha_1)(m+ \alpha_1)}\ ,\qquad \Delta F\equiv F_{\rm i}-F_{\rm ii}\ .
\end{align}
Therefore the transition is, as in the first analytic continuation, third order.
It occurs for generic, positive values of $ \alpha_1,\  \alpha_2$ at $ \alpha_1+ \alpha_2=m$.
Note that this phase transition is not meaningful in the ABJM case: it occurs when $ \alpha_1+ \alpha_2>m$,
which is never the case in ABJM where $ \alpha_1+ \alpha_2=0$.

\newpage


\section{Mass/FI-deformed ABJM theory does not have phase transitions \label{abjmtwo}}

In previous sections we solved the mass-deformed ABJM model with a non-vanishing FI term at large $N$, with the couplings analytically continued into the complex plane.
This is a standard approach in studying the large $N$ behaviour in Chern-Simons theories (see, e.g., \cite{Marino:2011nm,Drukker:2010nc,Suyama:2011yz,Suyama:2012uu}).
We have found a rich structure of phase transitions, but an important question is whether the  ABJM model with physical couplings and masses 
exhibits phase transitions as the coupling is varied.
For the mass-deformed models, analytic continuation back to physical couplings is not straightforward, due to the existence of poles originating from  the hyperbolic functions in the partition function. 
Clearly, it would be more desirable to have a direct solution of the large $N$ ABJM model with
the original parameters, with  $k_1=-k_2$ integers. 
Solving the saddle-point equations in this case turns out to be very complicated, 
 because eigenvalues seem to be distributed in cuts in the complex plane, with non-uniform $N$ dependence.
In this section we will argue that physical ABJM theory deformed by arbitrary mass and FI terms
is free from phase transitions.

\subsection{General case}

We start with  \eqref{eq:matrix_model_FI_shifted}, with  $m \leq 2\zeta $.
In section \ref{rtu} we have shown that this theory does not have  phase transitions for 
generic $\lambda_1,\ \lambda_2$, real and positive. 
Therefore the free energy
is a smooth function of the couplings $\lambda_1,\lambda_2$, indicating that there
should not be phase transitions after $\lambda_{1,2}\to e^{\pm i\varphi} \lambda_{1,2}$ analytic continuation back to the physical ABJM model when $m \leq 2\zeta $.
But the partition function of the ABJM model $Z_{\rm ABJM} (2\zeta,m)$ (with $N_1=N_2$) has the symmetry \eqref{fich}. The symmetry holds for any $N$, in particular, in  the planar limit $N\to\infty $ with fixed $k/N$.
If for given $\zeta $ and $m$, $Z_{\rm ABJM} (2\zeta,m)$ is a smooth function of $k/N$,
then so is $Z_{\rm ABJM} (m,2\zeta)$.
This indicates that there should be no phase transitions in ABJM neither in the opposite regime when  $m > 2\zeta $.

\subsection{Massive $U(2)\times U(2)$ ABJM \label{Utwo}}


We now consider a case where the partition function can be computed exactly, namely the $N=2$ case, i.e. 
$U(2)\times U(2)$ ABJM deformed by mass and  FI terms.
We will  first consider   the analytically continued model with real, equal couplings, particularised at $N=2$. 
This  corresponds to the second analytic continuation in the Chern-Simons levels, defined in section \ref{remak} and further
studied in section \ref{segunda}
(now the first analytic continuation cannot be used because the ranks of the two gauge groups are fixed
from the beginning).
We will show that, in the same decompactification limit, the model also exhibits phase transitions
of the same nature as the large $N$ model. 
Then we will discuss the $U(2)\times U(2)$  ABJM with physical coupling and parameter deformations, and
 show that there is no phase transition in this case.

\subsubsection{ The analytically continued model with equal, real couplings}

Consider the gauge group  $U(2)\times U(2)$, and analytic continuation in the  Chern-Simon levels to equal, real couplings. 
The partition function (\ref{eq:matrix_model_FI}) takes the form
\begin{align}
\label{tres}
Z = \frac{1}{4} \int \frac{d^2\mu}{(2\pi)^2} \frac{d^2 \nu}{(2\pi)^2} \frac{ \sinh^2 \frac{R}{2}(\mu_1-\mu_2) \sinh^2\frac{R}{2} (\nu_1-\nu_2)  \ e^{-\frac{R^2}{2g}\sum_i (\mu_i^2+  \nu^2_i) }}{\prod_{i,j=1}^2\cosh(\frac{R}{2}(\mu_i-\nu_j+m))\cosh(\frac{R}{2}(\mu_i-\nu_j-m))}
\end{align}
where  the $R$-dependence has been restored. In this subsection, for simplicity we set the FI-parameter to zero, since this case already illustrates the main point.
This is the analog of the models with $\alpha_1=\alpha_2$ studied at large $N$ 
 in section 4 of \cite{Anderson:2014hxa}), and briefly reviewed in section \ref{remak}. They have
similar phase transitions as the $\lambda_1=\lambda_2$ models obtained by the  analytic continuation
in the gauge group ranks, discussed in detail in  sections 4,5, 6.

Although phase transitions in matrix models typically arise at large $N$,
this type of phase transitions due to the contribution of extra massless multiplets at certain couplings  have also shown up in some  finite $N$ examples \cite{Russo:2014bda,Russo:2014nka}. The reason is that the contribution of an extra massless multiplet produces a singular behaviour at any $N$, even when $N=2$.
We now wish to see if the present model also has phase transitions.

The non-analytic behaviour arises upon taking a suitable decompactification limit $R\to \infty $, where
the  integral defining the partition function is dominated
by large  expectation values $\mu_i,\ \nu_i$.
In this limit, the hyperbolic cosine functions in the denominator get replaced by the non-analytic functions $\frac{1}{2} \exp|\mu_i-\nu_j\pm m|$, which produce non-analytic behaviour when the coupling is such that
$\mu_i-\nu_j$ at  the saddle-point hit $\pm m$.
We therefore assume the same scaling we used in the large $N$ case:
\begin{align}
g \equiv \lambda R \ ,
\end{align}
$$
R\to\infty\ ,\qquad \lambda \ {\rm fixed}.
$$
For large $R$, $Z$ may be written as:
\begin{align}
Z = 4\int \frac{d^2\mu}{(2\pi)^2} \frac{d^2 \nu}{(2\pi)^2}  e^{ R\big[ |\mu_1-\mu_2|+ |\nu_1-\nu_2|
-\frac{1}{2} \sum_{i,j=1}^2 (|\mu_i-\nu_j-m|+|\mu_i-\nu_j+m|)- \frac{1}{2\lambda}\sum_i (\mu_i^2+ \nu^2_i)\big]} 
\end{align}
and the integral is dominated by a saddle-point. The saddle-point equations are
\begin{align}
\frac{1}{\lambda} \;  \mu_1 & =   {\rm sign}(\mu_1-\mu_2)\;  - \frac{1}{2} \; \sum_{j=1}^2 \big[{\rm sign}(\mu_1-\nu_j-m) +{\rm sign}(\mu_1-\nu_j+m) \big]
\noindent\\
\frac{1}{\lambda} \;  \mu_2 & =  -{\rm sign}(\mu_1-\mu_2) \; - \frac{1}{2} \; \sum_{j=1}^2 \big[{\rm sign}(\mu_2-\nu_j-m) +{\rm sign}(\mu_2-\nu_j+m) \big]
\noindent\\
\frac{1}{\lambda} \;  \nu_1 & =   {\rm sign}(\nu_1-\nu_2)\;  - \frac{1}{2} \; \sum_{j=1}^2 \big[{\rm sign}(\nu_1-\mu_j-m) +{\rm sign}(\nu_1-\mu_j+m) \big]
\noindent\\
\frac{1}{\lambda} \;  \nu_2& =  -{\rm sign}(\nu_1-\nu_2)\;  - \frac{1}{2} \; \sum_{j=1}^2 \big[{\rm sign}(\nu_2-\mu_j-m) +{\rm sign}(\nu_2-\mu_j+m) \big]
\end{align}
We can assume with no loss of generality $\mu_1>\mu_2$ and $\nu_1>\nu_2$.
If the eigenvalues are sufficiently small, their difference will be less than $m$.
Then the sign functions containing $m$ in the argument cancel out and we find the solution
\begin{align}
\mu_1= \lambda ,\quad \mu_2=-\lambda\ ,\quad \nu_1=\lambda\ ,\quad \nu_2=-\lambda\ .
\end{align}
Thus this solution holds for\footnote{For two independent couplings $\lambda_1,\ \lambda_2$, the solution is
$\mu_1=-\mu_2=\lambda_1$, $\nu_1=-\nu_2=\lambda_2$, with $\lambda_1+\lambda_2<m$,
which is the $N=2$ analog of the solution \eqref{onepa}.}
\begin{align}
\lambda< \frac{m}{2}\ .
\end{align}
When $\lambda >\frac{m}{2}$, the solution will change, because the difference of eigenvalues can overcome $m$ and in this case the sign functions will contribute. 
In this $\lambda >\frac{m}{2}$ regime, the absolute minimum of the action is given by
\begin{align}
\mu_1= \nu_1= \frac{m}{2}  ,\quad \mu_2= \nu_2=-\frac{m}{2}\ .
\end{align}
In this case the arguments of some sign functions vanish. The action is not differentiable at this point
and the minimum must be found by inspection.

Therefore the theory contains two phases.
The free energy $F=-\frac{1}{R} \ln Z$ in each phase is given by the action evaluated at the minimum
of the potential. We find
\begin{align}
F= \begin{cases}
  4m-2\lambda  \hspace*{6mm} \lambda< \frac{m}{2}
 \\ \nonumber
 2m+\frac{m^2}{2 \lambda} \hspace*{6mm} \lambda \geq \frac{m}{2}
 \end{cases}
\end{align}
This implies a discontinuity in the second derivative
\begin{align}
\Delta F  \Big|_{\lambda= \frac{m}{2}  } = \partial_\lambda \Delta F \Big|_{\lambda= \frac{m}{2}  } = 0\ ,\qquad  \partial^2_\lambda \Delta F \Big|_{\lambda= \frac{m}{2}  } = -\frac{8}{m}\ .
\end{align}
Therefore, we conclude that the analytically continued $U(2)\times U(2)$ mass-deformed ABJM model presents phase transitions.


\subsubsection{$U(2)\times U(2)$ ABJM model with physical couplings}

 The partition function of the $U(2)\times U(2)$ ABJM deformed by mass and a FI term
computed by localisation is given (see (\ref{eq:matrix_model_FI}) with $N=2$)
\begin{align}
\label{unamas}
Z = \frac{1}{4} \int \frac{d^2\mu}{(2\pi)^2}\ \frac{d^2 \nu}{(2\pi)^2} \frac{ \sinh^2 \frac{\mu_1-\mu_2}{2} \sinh^2\frac{\nu_1-\nu_2}{2}}{\prod_{i,j=1}^2\cosh(\frac{\mu_i-\nu_j+m}{2})\cosh(\frac{\mu_i-\nu_j-m}{2})}
 \ e^{\frac{i k}{4\pi}\sum_i (\mu_i^2 -   \nu^2_i)-  \frac{ik}{2\pi} \zeta  (\sum_i \mu_i+\sum_i \nu_a ) }
\end{align}
Using  \eqref{permuform} for $N=2$, the partition function can be written in the following form 
\begin{align}
Z = \frac{1}{2}\left( Z_1- Z_2 \right)\ ,
\end{align}
with
\begin{align}
Z_1 = \int d\tau_1 d\tau_2 \  \frac{ e^{- i k m_2(\tau_1+\tau_2)} }{\cosh (\pi k\tau_1)
\cosh (\pi k \tau_2) \cosh^2 \big(\frac{m_1}{2} \big)} \ ,
\end{align}
and 
\begin{align}
\label{gar}
Z_2=
\int d\tau_1 d\tau_2 \  \frac{ e^{- i k m_2(\tau_1+\tau_2)} }{\cosh (\pi k\tau_1)
\cosh (\pi k \tau_2)
\cosh \big(\pi (\tau_1-\tau_2)- \frac{m_1}{2})\big)\cosh \big(\pi (\tau_1-\tau_2)+\frac{m_1}{2})\big)}   \ ,
\end{align}
\begin{align}
m_1 \equiv m+2\zeta   \ ,\qquad m_2 \equiv m- 2\zeta \ .
\end{align}
It is important to note that the derivation that leads to this form of the partition function 
holds if and only if the Chern-Simons levels of both gauge groups in
 $U(2)_{k_1}\times U(2)_{k_2}$ are opposite, i.e. $k_2=-k_1$.

The first integral can be computed by using the Fourier integral \eqref{fouri}.
We get
\begin{align}
Z_1=\frac{1} {k^2\cosh^2\big(\frac{m_1}{2}\big)\cosh^2 \big(\frac{m_2}{2}\big)}\ .
\end{align}
Next, consider the calculation of $Z_2$. Note that we have reduced the original four integrals to only two integrals $\tau_1$ and
$\tau_2$.
The integral (\ref{gar}) can be carried out by defining variables $u=\tau_1+\tau_2$, $v=\tau_1-\tau_2$.
Then the integral over $u$ is a Fourier transform than can be computed explicitly:
\begin{align}
\label{fou2}
\int du  \frac{e^{- i k m_2 u}}{\cosh \frac{\pi k}{2} (u+v) \cosh \frac{\pi k}{2} (u-v)} = \frac{2\sin ( k  m_2 v) }{k \sinh(\pi k v) \sinh m_2 }\ .
\end{align}
Hence
\begin{align}
Z_2 =\frac{2}{k \sinh  m_2} \int dv  \frac{\sin( k m_2  v)}{\sinh(\pi  k v) \cosh(\pi v-\frac{m_1}{2})\cosh(\pi v+\frac{m_1}{2})}\ .
\end{align}
Restoring the $R$ dependence, and rescaling $v\to R v$, we have
\begin{align}
Z_2 =\frac{2\lambda R^2}{\sinh( m_2 R)}   \int dv  \frac{\sin( m_2  v R/\lambda)}{\sinh(\pi   v/\lambda) \cosh( R(\pi v-\frac{m_1}{2}))\cosh(R( \pi v+\frac{m_1}{2}))}
\end{align}
where we defined $\lambda =1/(kR)={\rm fixed}$.
At large $R$, this integral is not dominated by a saddle-point; nevertheless, it can be computed exactly.
For $R\to\infty $, the product of the hyperbolic cosines  in the denominator becomes proportional to a step-function with support in the
interval $(- \frac{m_1}{2\pi},\frac{m_1}{2\pi})$. The $\sin(\pi v/\lambda )$ in the denominator can be replaced by $\pi v/\lambda$
(as can be seen e.g. by a change of integration variable $v\to x/R$).
The resulting integral can be carried out explicitly, with the result
\begin{align}
Z_2 
=\frac{32}{\pi k^2}   e^{- R( |m_1|+|m_2|)} \, {\rm Si}( \frac{k }{2\pi} |m_1m_2| R^2 )\ ,
\end{align}
where ${\rm Si}(z)$ is the Sine integral function.
Combining with $Z_1$ at large $R$, we finally obtain
\begin{align}
Z &= \ 
\frac{8}{k^2}   e^{- R( |m_1|+|m_2|)} \left(1-\frac{2}{\pi} {\rm Si}( \frac{k }{2\pi} |m_1m_2| R^2 )\right) 
\nonumber\\
&\approx  \ 
 \frac{32}{ k^3 |m_1m_2| R^2}   e^{- R( |m_1|+|m_2|)} \cos(\frac{k }{2\pi} |m_1m_2|R^2 )\ ,
\label{finalZ}
\end{align}
where we have used the asymptotic expansion of the Sine integral function.

Thus we have evaluated the $U(2)_k\times U(2)_{-k}$ ABJM partition function with both mass and FI-parameter deformations in the large $R$ limit.
In particular,  this shows that $F= -\frac{1}{R} \ln Z $ is an analytic function of the coupling $k$ (or $\lambda $).
Therefore, for physical couplings, the model does not exhibit phase transitions.
More generally, since the Sine integral function is an entire function in the whole complex plane,
the theory does not have phase transition in any region of the $k$-complex plane.
Thus, phase transitions seem to be absent in  $U(2)\times U(2)$ ABJM theory, where the Chern-Simons levels of the gauge group are  opposite integers. 

\bigskip

\noindent {\bf Remark:} The partition function \eqref{finalZ} has zeroes, which, at large R, are located  at
\begin{align}
km_1m_2 R^2= \frac{m_1m_2 R}{\lambda} \approx \pi^2    (2n+1)\ ,\quad n\in {\mathbb Z}
\end{align}
These are Lee-Yang singularities.\footnote{We thank K. Zarembo for this remark.} It is easy to see that the partition function has  zeroes also at finite $R$.
It would be interesting to get further insights on their physical meaning. They appear to be resonances occurring at special values of $mR$, perhaps associated with Kaluza-Klein excitations in ${\mathbb S}^3$.
They seem to arise by virtue of the fact that: a) Chern-Simons theory has imaginary coupling $g=2\pi i/k$; 
b) the theory has mass/FI deformation; c) the theory is on a compact space. 
As a small check, one can see that the partition function of other Chern-Simons theories with massive matter also exhibit similar zeroes. In particular, for
$U(2) $ Chern-Simons theory with fundamental matter one finds \cite{Russo:2014bda}
\begin{align}
Z^{U(2)}_{(k=2)} = \frac{8\pi^2 e^{2mR} \left( e^{\frac{im^2R^2}{2\pi}}-1\right)\left( e^{\frac{im^2R^2}{2\pi}}+i\right)}{\left(e^{2mR}-1\right)^2}\ ,
\end{align}
which, indeed, has an infinite number of zeroes at $m^2R^2=  \pi^2 n $, $m^2R^2=  \pi^2 (4n-1) $, $n=1,2,...$

\section{Concluding remarks \label{concluding}}

In the first part of this paper we have studied the general solution to functional equations of the form
\begin{align}
\rho(z) =& \frac{1}{2 \lambda}-
 \frac{1}{2} \rho(-z-m_1)
- \frac{1}{2} \rho(-z+m_2)
.\end{align}
where $\rho(z)$ is a unit-normalised density supported  on some interval $[-A,B]$ along the real axis. 
We have shown that this equation describes the large $N$ limit of the mass- and FI-deformed 
ABJ partition function $Z(\lambda_1,\lambda_2)$ 
with couplings $\lambda_1=2\pi i N_1/k$, $\lambda_2=2\pi i N_2/k$ analytically continued to the complex
plane, in the particular region where $\lambda_1=\lambda_2 \in \mathbb{R}$.
The study herein generalises the discussion of \cite{Anderson:2014hxa}, corresponding to the case $m_1=m_2$ in this notation, and this generalisation turns out to be highly non-trivial since it amounts to giving up the reflection symmetry around the origin which drastically simplified the analysis in the special case of $m_1=m_2$.

Unlike the case of vanishing FI-parameter, for $\zeta\neq 0$  the theory exhibits a finite number of phases
 as the coupling is increased from 0 to infinity.
The structure of these phases is summarised in section \ref{sec:Conclusion} and in the complete phase diagram  given in figure \ref{fig:phase_diagram}. One result obtained here of particular interest is that the FI deformation alone, while it introduces a mass scale, does not generate phase transitions.
Furthermore, in section \ref{sec:Conclusion}, we have also considered the cases of generic (real and positive) couplings $\lambda_1,\ \lambda_2$, 
as well as the case of analytic continuation in the  Chern-Simons levels instead, and showed that the qualitative 
picture is similar to the equal coupling case of previous sections.

An important question concerns the implications of these results for the physical ABJM theory,
where the couplings are opposite ($\lambda_1=-\lambda_2 $) and purely imaginary. 
We have presented an argument showing that there are no phase transitions in this case.
The argument relies on the symmetry of the original ABJM partition function under exchange of
FI- and mass-deformations, along  with the fact, derived in section \ref{rtu}, that the analytically continued theory is free from phase transitions when $\zeta>m/2$. 
To further clarify this issue,
we have computed the partition function for the gauge group $U(2)\times U(2)$ exactly. While  the theory has phase transitions
of the same nature as the large $N$ model in a region of parameter space (in particular, at equal real, couplings), there are nevertheless no phase transitions in the physical ABJM case, where the 
Chern-Simons levels are $k$ and $-k$.

The $U(2)_k\times U(2)_{-k}$ example explains why  the $U(N)_k\times U(N)_{-k}$ ABJM model should be free from  phase transitions. It would be interesting to get further insights on the whole picture and on the analytic structure of the mass/FI-deformed ABJM partition function.

\bigskip
\bigskip

\section*{Acknowledgement}
We are grateful to K. Zarembo for  useful discussions. J.R. acknowledges financial support from projects  FPA2013-46570  and
 2014-SGR-104.

\newpage
\appendix

\section{The phases for $m_2>B-A$ \label{app}}

In section \ref{sec:mtilde_geq_B-A} we found that,
when $m_2>B-A$,   even and odd resonances can get into the interval $[-A,B]$ of support of the density in three different ways:

\begin{itemize}

\item  {\bf Case I:} Both points $a_{2n+1}$ and $b_{2n+1}$ lie outside the interval, which is equivalent to 
\begin{align}
 \frac{A-\zeta}{m}-\frac{1}{2} \;<\;   \frac{B+\zeta}{m}-\frac{1}{2}  \; <\;   n.
 \end{align}
\item {\bf Case II:} The point $a_{2n+1}$ lies outside the interval, but $b_{2n+1}$ lies inside the interval, corresponding to 
\begin{align}
 \frac{A-\zeta}{m}-\frac{1}{2} \;<\; n \; <\; \frac{B+\zeta}{m}-\frac{1}{2} 
 .\end{align}
\item {\bf Case III:} Both $a_{2n+1}$ and $b_{2n+1}$ lie inside the interval, that is,   
\begin{align}
  n \; <\; \frac{A-\zeta}{m}-\frac{1}{2} \;<\; \frac{B+\zeta}{m}-\frac{1}{2} 
.\end{align}
\end{itemize}

\noindent In what follows we examine each case separately.

\subsection{Case I:  $n =\left[ \frac{A+B}{2m} \right], \quad  \frac{A-\zeta}{m}-\frac{1}{2} <  \frac{B+\zeta}{m}-\frac{1}{2} <  n $. \label{sec:case_1}}

\begin{figure}[t]
\begin{center}
 \centerline{\includegraphics[width=11cm]{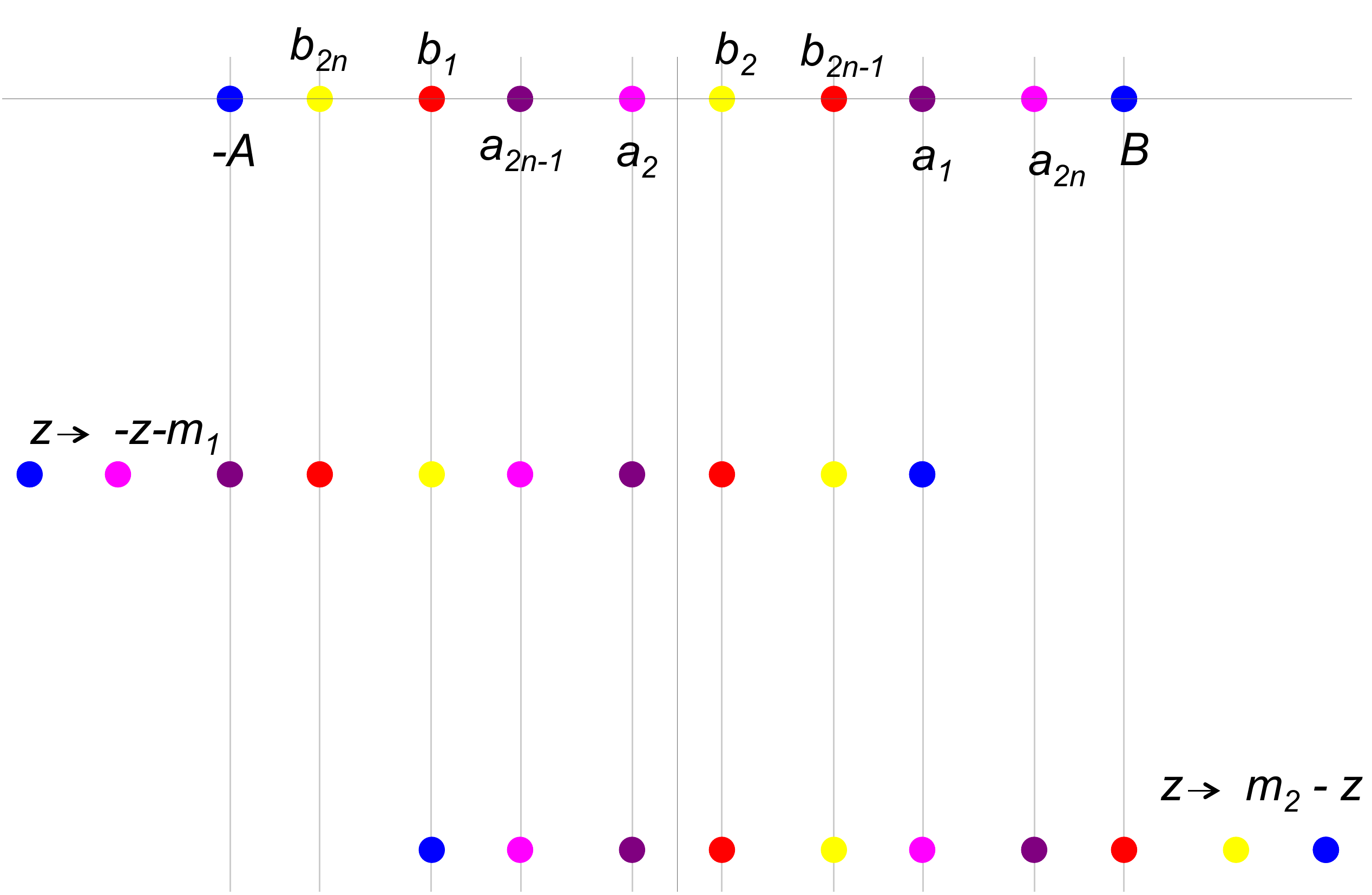}}
\caption{\label{fig:SmallFI_int_res_from_AB_EvenN}\small  Resonance points originating from the interval endpoint $-A$ and $B$ (which are shown in blue). Resonances originating from $-A$ are shown in  magenta (even, $a_{2k}$), and  purple (odd resonances, $a_{2k-1}$). Similarly, resonances originating from the rightmost interval endpoint, $B$, are shown in  yellow (even, $b_{2k}$) and  red (odd, $b_{2k-1}$). These resonances exist for all $k \leq \frac{B+A}{2m}$. Furthermore, the resonances under the maps $z \rightarrow -z-m_1$ and $z \rightarrow m_2-z$ are shown underneath. }
\end{center}
\end{figure}

In this case, the ordering of the resonance points in the interior of the interval will be given by: \begin{align}
-A & \; < \;  b_{2n}\; < \;  b_1 \; < \;  a_{2n-1} \; < \; 
\\ \nonumber &
\; < a_2 \; < \;  b_{2(n-1)} \; < \; b_{3}  \;< \; a_{2(n-1)-1} \; < \; \dots \; <
 \\ \nonumber &
\quad \quad \;\;
<a_{2n-2} \; <\;  b_2 \; <\; b_{2n-1} \; < \;a_1 \; < \;a_{2n} < B
,\end{align}
which is illustrated in figure \ref{fig:SmallFI_int_res_from_AB_EvenN}. With a piecewise constant Ansatz for the density on the form:
\begin{align}
\rho(z)=&
\begin{cases}
\alpha\beta_k 
\qquad
  \forall z \in [a_{2k},b_{2(n-k)}]
&
k \in [0 , n]
\\
\beta_k  
\qquad  \;\;\,
 \forall z \in [b_{2(n+1-k)},b_{2k-1}]
&
k \in [1, n]
 \\
\beta\alpha_k  
\qquad 
\forall z \in [b_{2k-1},a_{2(n-k)+1}]
&
k\in [1,n]
\\
\alpha_k 
\qquad  \;\;\,
 \forall z \in [a_{2(n-k)+1},a_{2k}],  
&
 k \in [1 , n]
\end{cases}
,\end{align}  
with $-A=a_0, B=b_0$, the saddle-point equation \eqref{eq:saddles_ana1_equal} 
for $\alpha\beta_k,\beta_k,\beta\alpha_k,\alpha_k$ takes the form:
\begin{align}
2\alpha\beta_k+
\beta\alpha_{n-k}
+
\beta\alpha_{n+1-k}
 \; =& \;
\frac{1}{\lambda} & k \in [0,n]
\\ \nonumber
2\beta_k+
\beta_{n+1-k}
+
\beta_{n+2-k}
 \; =& \;
\frac{1}{\lambda} & k \in [1,n]
\\ \nonumber
2\beta\alpha_k+
\alpha\beta_{n-k}
+
\alpha\beta_{n+1-k}
 \; =& \;
\frac{1}{\lambda} & k \in [1,n]
\\ \nonumber
2\alpha_k+
\alpha_{n-k}
+
\alpha_{n+1-k}
 \; =& \;
\frac{1}{\lambda} & k \in [1,n]
,\end{align}
with the boundary conditions $\beta_{n+1}=\alpha_0=\beta\alpha_{n+1}=\beta\alpha_{0}=0$.
The decoupled equations, for $\alpha, \beta$, are straightforward to solve, and from the two remaining coupled equations, one finds that the general solution for $\beta\alpha_k$ is a linear function in $k$. Enforcing the boundary conditions, both integration constants vanish,  completely specifying $\beta\alpha_k$, and thus also $\alpha\beta_k$.
Altogether, one then finds the eigenvalue density as:

\begin{align}
\rho(z)=&
\begin{cases}
\alpha\beta_k  
\;
=\frac{1}{2\lambda} 
\qquad \quad \;\;\;
  \forall z \in [a_{2k},b_{2(n-k)}]
&
k \in [0 , n]
\\
\beta_k  
\;  \;\;\, 
= \frac{1+n-k}{\lambda(1+2n)} 
\qquad
 \forall z \in [b_{2(n+1-k)},b_{2k-1}]
&
k \in [1, n]
 \\
\beta\alpha_k   
\;
=0
 \qquad \qquad \;
\forall z \in [b_{2k-1},a_{2(n-k)+1}]
&
k\in [1,n]
\\
\alpha_k
\;  \;\;\,
= \frac{k}{\lambda(1+2n)}
\qquad
 \forall z \in [a_{2(n-k)+1},a_{2k}],  
&
 k \in [1 , n]
\end{cases}
,\end{align}
which simplifies to the solution found previously in \cite{Anderson:2014hxa} for the case of a vanishing FI-parameter.

\subsection{Case II: $n =\left[ \frac{A+B}{2m} \right], \quad \frac{A-\zeta}{m}-\frac{1}{2} < n <  \frac{B+\zeta}{m}-\frac{1}{2}$. \label{sec:case_2}}
In this case, the point $b_{2n+1}$ lies inside the interval $[-A,B]$, but the point $a_{2n+1}$ does not. The ordering of the interior resonance points in this case will be as in the previous, but with odd- and even $b$-resonances interchanged, that is: 
\begin{align}
-A & \; < \; b_1  \; < \;    b_{2n}\; < \;   a_{2n-1} \; < \; 
\\ \nonumber &
\; < a_2 \; < \;  b_{3}  \;< \;  b_{2(n-1)} \; < \;  a_{2(n-1)-1} \; < \; \dots \; <
 \\ \nonumber &
\quad \quad \;\;
<a_{2n-2} \; <\; b_{2n-1} \; < \;  b_2 \; <\;  a_1 \; < \;a_{2n} \; < \; b_{2n+1} \; < B
.\end{align}

Again, the density will be given by a piecewise constant one, which, using the notations $\alpha \beta_k,\beta_k, \beta \alpha_k, \alpha_k$ are governed by the following equations:
\begin{align}
2\alpha\beta_k+
\beta\alpha_{n-k}
+
\beta\alpha_{n+1-k}
 \; =& \;
\frac{1}{\lambda} & k \in [0,n]
\\ \nonumber
2\beta_k+
\beta_{n-1-k}
+
\beta_{n-k}
 \; =& \;
\frac{1}{\lambda} & k \in [0,n]
\\ \nonumber
2\beta\alpha_k+
\alpha\beta_{n-k}
+
\alpha\beta_{n+1-k}
 \; =& \;
\frac{1}{\lambda} & k \in [1,n]
\\ \nonumber
2\alpha_k+
\alpha_{n-k}
+
\alpha_{n+1-k}
 \; =& \;
\frac{1}{\lambda} & k \in [1,n]
,\end{align}
together with the boundary conditions
\begin{align}
\alpha_0=\beta_{-1}=\beta\alpha_0=\beta\alpha_{n+1}=0
.\end{align}

This gives us the eigenvalue density as 
\begin{align}
\rho(z)=&
\begin{cases}
\alpha\beta_k  
\;
=\frac{1}{2\lambda} 
\qquad \quad \;\;\;
  \forall z \in [a_{2k},b_{2k+1} ]
&
k \in [0 , n]
\\
\beta_k  
\;  \;\;\, 
= \frac{1+k}{\lambda(3+2n)} 
\qquad
 \forall z \in [b_{2k+1},b_{2(n-k)}]
&
k \in [0, n]
 \\
\beta\alpha_k   
\;
=0
 \qquad \qquad \;
\forall z \in [b_{2(n-k+1)},a_{2(n-k)+1}]
&
k\in [1,n]
\\
\alpha_k
\;  \;\;\,
= \frac{k}{\lambda(1+2n)}
\qquad
 \forall z \in [a_{2(n-k)+1},a_{2k}],  
&
 k \in [1 , n]
\end{cases}
.\end{align}

\subsection{ Case III: $n =\left[ \frac{A+B}{2m} \right], \quad  n < \frac{A-\zeta}{m}-\frac{1}{2} < \frac{B+\zeta}{m}-\frac{1}{2}$. \label{sec:case_3}}

\begin{figure}[t]
\begin{center}
 \centerline{\includegraphics[width=11cm]{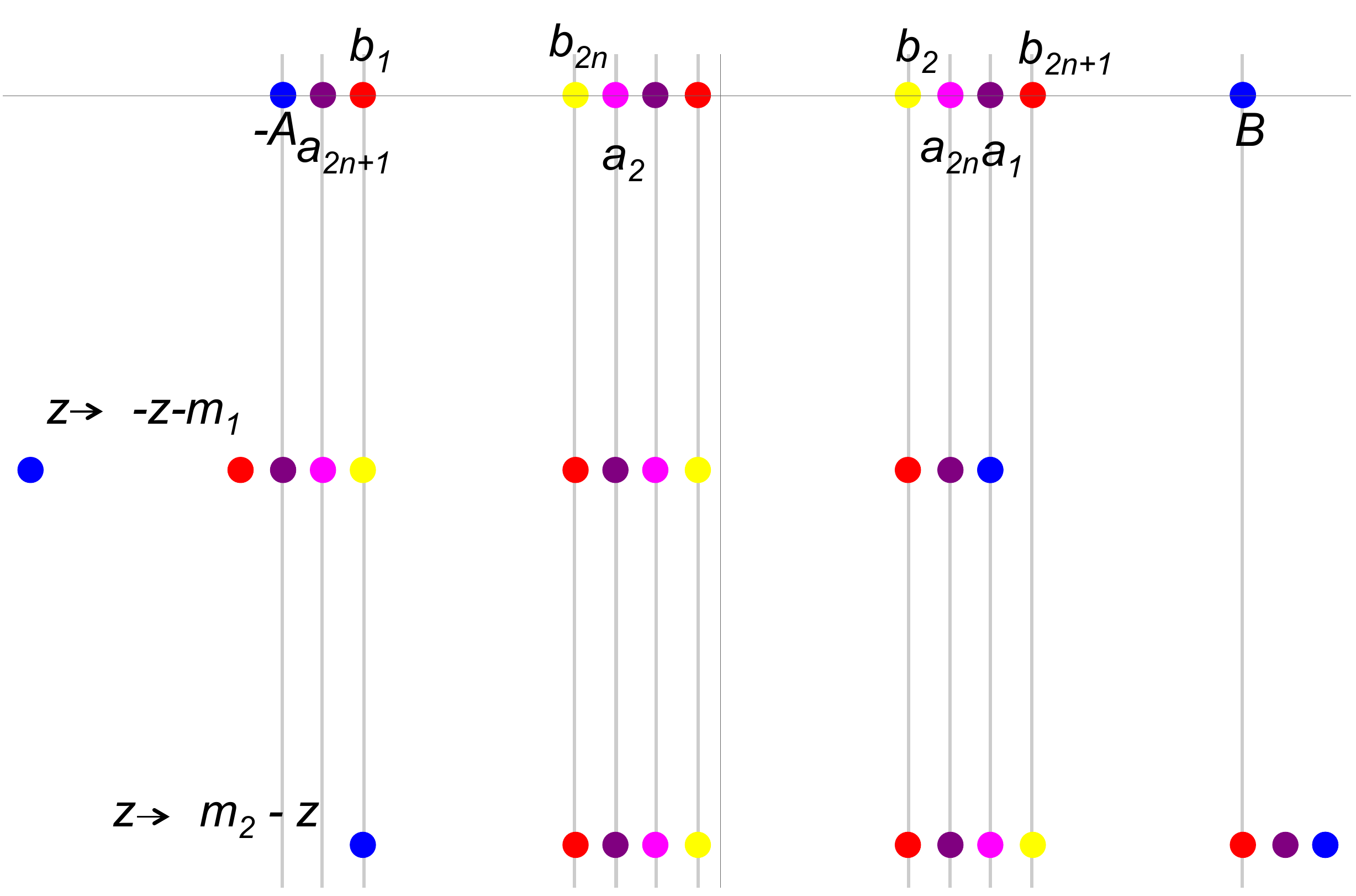}}
\caption{\label{fig:SmallFI_int_res_from_AB_case3}\small  Resonance points originating from the interval endpoint $-A$ and $B$ (which are shown in  blue). Resonances originating from $-A$ are shown in  magenta (even, $a_{2k}$), and  purple (odd resonances, $a_{2k-1}$). Similarly, resonances originating from the rightmost interval endpoint, $B$, are shown in yellow (even, $b_{2k}$) and  red (odd, $b_{2k+1}$). These resonances exist for all $k \leq \frac{A+B}{2m}$. Furthermore, the resonances under the maps $z \rightarrow -z-m_1$ and $z \rightarrow m_2-z$ are shown underneath. }
\end{center}
\end{figure}

In this case, both $a_{2n+1}$ and $b_{2n+1}$ will lie in the interior of the interval, and the ordering amongst the interior resonance points will be as in the second case, but with interchanged ordering amongst the $a_k$'s. Furthermore, since the point $a_{2n+1}$ now enters the interior of the interval, this will be the leftmost resonance point, and not $b_1$ as in the previous case. The ordering may be seen in figure \ref{fig:SmallFI_int_res_from_AB_case3}.

\vspace*{1cm}

Once more, we write down a piecewise constant Ansatz for the density, the only change from the previous case is the numbering and precise appearance on the boundary conditions for $\alpha_k, \beta_k, \alpha\beta_k, \beta\alpha_k$. Let our Ansatz be:
\begin{align}
\rho(z) =
\begin{cases}
\alpha_k 
& \forall z \in [a_{2k},a_{2(n-k)+1}],  
\hspace*{2.5cm} 
 k \in [0 , n]
\\
\alpha\beta_k 
&  \forall z \in [a_{2(n-k)+1},b_{2k+1}]
\hspace*{2.4cm} 
k \in [0 , n]
\\
\beta_k  
& \forall z \in [b_{2k+1},b_{2(n-k)}]
\hspace*{2.8cm}  k \in [0, n]
 \\
\beta\alpha_k  
&
\forall z \in [b_{2(n+1-k)},a_{2k}]
\hspace*{2.8cm} 
k\in [1,n]
\end{cases}
.\end{align}
Then, the saddle-point equation \eqref{eq:saddles_ana1_equal} takes the form:
\begin{align}
2\alpha_k + \alpha_{n-k}
+\alpha_{n+1-k}
=& 
\frac{1}{\lambda} 
& k \in [0,n]
\\ \nonumber
2\alpha\beta_k 
+\beta\alpha_{n-k}
+ \beta\alpha_{n+1-k}
=& 
\frac{1}{\lambda} 
& k \in [0,n]
\\ \nonumber
2\beta_k 
+ \beta_{n-1-k}
+ \beta_{n-k}
=& 
\frac{1}{ \lambda} 
& k \in [0,n]
\\ \nonumber
2\beta\alpha_k 
+\alpha\beta_{n-k}
+ \alpha\beta_{n+1-k}
=& 
\frac{1}{\lambda} 
& k \in [1,n]
,\end{align}
with
\begin{align}
\alpha_{2n+1}=\beta_{-1}= \beta\alpha_{0}= \beta\alpha_{n+1}=0
.\end{align}

Hence, in general, the eigenvalue density in this case is given by:
\begin{align}
\label{eq:density_case3}
\rho(z) =
\begin{cases}
\alpha_k \;\; =\frac{n+1-k}{ \lambda (3+2  n)}
& \forall z \in [a_{2k},a_{2(n-k)+1}],  
\hspace*{2.5cm} 
 k \in [0 , n]
\\
\alpha\beta_k \; =\frac{1}{2\lambda}
&  \forall z \in [a_{2(n-k)+1},b_{2k+1}]
\hspace*{2.4cm} 
k \in [0 , n]
\\
\beta_k  \;\; \;= \frac{k+1}{\lambda (3+2  n)}
& \forall z \in [b_{2k+1},b_{2(n-k)}]
\hspace*{2.8cm}  k \in [0, n]
 \\
\beta\alpha_k  \;=0
&
\forall z \in [b_{2(n+1-k)},a_{2k}]
\hspace*{2.8cm} 
k\in [1,n]
\end{cases}
,\end{align}
which indeed reduces to the solution previously found for vanishing FI-parameter as $B \rightarrow A$ and $ \zeta \rightarrow 0$.

\subsection{Determining the interval endpoints}

Once more, the rightmost interval endpoint may be determined from the integral equation \eqref{eq:saddlepoints_cont_rescaled}, where the two first integrals are  simply determined from the normalisation condition. However, the third one is not.  The $\sign$-function in that integral will take the value one for $y >m_2-B=b_1$, whereas it will take the value $-1$ for $y< m_2-B=b_1$, and so we find $B$ as:

\begin{align}
B=
 & \frac{\lambda}{2} \; \Big( \;
 3  
-    \int_{-A}^{b_1} d y\,\rho  (y )
+    \int_{b_1}^{B} d y\,\rho  (y )
\Big)
.\end{align}
These  integrals will depend on the ordering on the resonance points and thus will take different values in the different cases. 
Define the distances between one $a$ and one $b$-resonance point as $\mathfrak{ab}$, the distance between two $a$-resonances as $\mathfrak{a}$, and between two $b$-resonances as $\mathfrak{b}$. The integral may then be divided as:

\begin{align}
B=
 & \frac{\lambda}{2}  \; \Big( \;
 3  
-   \mathfrak{ab} \sum_{<b_1} \alpha\beta_k 
-   \mathfrak{a} \sum_{<b_1}  \alpha_k 
-   \mathfrak{b} \sum_{<b_1} \beta _k 
\\ \nonumber &
+   \mathfrak{ab} \sum_{>b_1}  \alpha\beta_k 
+   \mathfrak{a} \sum_{>b_1}  \alpha_k 
+   \mathfrak{b} \sum_{>b_1} \beta _k 
\Big)
.\end{align}
Inserting the expressions for the eigenvalue densities, one then finds:
\begin{align}
\label{eq:small_FI_B_integral_eq}
B=  & \frac{\lambda}{2} \; \Big( \;
 3  
+\mathfrak{ab}\frac{ n-1}{2 \lambda}
\\ \nonumber &
-   \mathfrak{a} \sum_{<b_1}  \alpha_k 
-   \mathfrak{b} \sum_{<b_1} \beta _k 
+   \mathfrak{a} \sum_{>b_1}  \alpha_k 
+   \mathfrak{b} \sum_{>b_1} \beta _k 
\Big)
.\end{align}

Furthermore, the normalisation condition on the eigenvalue density will give us another condition on the interval endpoints. In the context of $\mathfrak{ab}, \mathfrak{a}$ and $\mathfrak{b}$, this takes the form:
\begin{align}
\label{eq:small_FI_normalisation}
\nonumber
\mathfrak{ab}\sum_{k}\alpha\beta_k+\mathfrak{a} \sum_k \alpha_k+\mathfrak{b} \sum_k \beta_k \; =& \; 1
\\ \Leftrightarrow \hspace*{2cm} 
\\ \nonumber
\mathfrak{ab}\frac{n+1}{2 \lambda}+\mathfrak{a} \sum_k \alpha_k+\mathfrak{b} \sum_k \beta_k \; =& \; 1
.\end{align}
Together, the equations \eqref{eq:small_FI_B_integral_eq} and \eqref{eq:small_FI_normalisation} gives us two relations between the interval endpoints and the parameters $m, \zeta$ and $\lambda$, and these may be used to determine the interval endpoints in terms of these quantities.

These expressions  depend on the precise appearance of the eigenvalue densities, as well as on the ordering of the resonance points around $b_1$. Therefore they will give rise to different results for the interval endpoints in the three cases. Inserting the appropriate limits in the sum, and the corresponding number of points to the left/right of $b_1$, one  finds  the interval endpoints  described in \eqref{casoI} -- \eqref{casoIII}.

\pagebreak

\bibliography{refs}{}
\bibliographystyle{JHEP}

\end{document}